\documentclass{emulateapj}

\def\dave{Dav{\'e}}

\begin{document}
\title{$z\sim7$ galaxy candidates from NICMOS observations over the
  HDF South and the CDF-S and HDF-N GOODS fields\altaffilmark{1}}
\author{Rychard J. Bouwens\altaffilmark{2,3}, Garth D. Illingworth\altaffilmark{2}, 
Valentino Gonzalez\altaffilmark{2}, Ivo Labb{\'e}\altaffilmark{4},
Marijn Franx\altaffilmark{3}, 
Christopher J. Conselice\altaffilmark{5}, John Blakeslee\altaffilmark{6}, 
Pieter van Dokkum\altaffilmark{7}, 
Brad Holden\altaffilmark{2}, Dan Magee\altaffilmark{2}, 
Danilo Marchesini\altaffilmark{9}, Wei Zheng\altaffilmark{8}}
\altaffiltext{1}{Based on observations made with the NASA/ESA Hubble
  Space Telescope, which is operated by the Association of
  Universities for Research in Astronomy, Inc., under NASA contract
  NAS 5-26555. These observations are associated with programs \#7235,
  7817, 9425, 9575, 9723, 9797, 9803, 9978, 9979, 10189, 10339, 10340,
  10403, 10530, 10632, 10872, 11082, 11144, 11192.  Observations have
  been carried out using the Very Large Telescope at the ESO Paranal
  Observatory under Program ID(s): LP168.A-0485.}
\altaffiltext{2}{Astronomy Department, University of California, Santa Cruz,
  CA 95064}
\altaffiltext{3}{Leiden Observatory, Leiden University, Postbus 9513,
  2300 RA Leiden, Netherlands}
\altaffiltext{4}{Carnegie Observatories, Pasadena, CA 91101, Hubble
  Fellow}
\altaffiltext{5}{School of Physics and Astronomy, University of
c  Nottingham, Nottingham, NG72RD, UK}
\altaffiltext{6}{Herzberg Institute of Astrophysics, Victoria, BCV9E2E7, 
Canada}
\altaffiltext{7}{Department of Astronomy, Yale University, New Haven, CT 06520}
\altaffiltext{8}{Department of Physics and Astronomy, Johns
  Hopkins University, 3400 North Charles Street, Baltimore, MD 21218}
\altaffiltext{9}{Department of Physics and Astronomy, Tufts University, Medford, MA 
02155}

\begin{abstract}
We use $\sim$88 arcmin$^2$ of deep ($\gtrsim$26.5 mag at $5\sigma$)
NICMOS data over the two GOODS fields and the HDF South to conduct a
search for bright $z\gtrsim7$ galaxy candidates.  This search takes
advantage of an efficient preselection over 58 arcmin$^2$ of NICMOS
$H_{160}$-band data where only plausible $z\gtrsim7$ candidates are
followed up with NICMOS $J_{110}$-band observations.  $\sim$248
arcmin$^2$ of deep ground-based near-infrared data ($\gtrsim25.5$ mag,
$5\sigma$) is also considered in the search.  In total, we report 15
$z_{850}$-dropout candidates over this area -- 7 of which are new to
these search fields.  Two possible $z\sim9$ $J_{110}$-dropout
candidates are also found, but seem unlikely to correspond to $z\sim9$
galaxies (given the estimated contamination levels).  The present
$z\sim9$ search is used to set upper limits on the prevalence of such
sources.  Rigorous testing is undertaken to establish the level of
contamination of our selections by photometric scatter, low mass
stars, supernovae (SNe), and spurious sources.  The estimated
contamination rate of our $z\sim7$ selection is $\sim$24\%.  Through
careful simulations, the effective volume available to our $z\gtrsim7$
selections is estimated and used to establish constraints on the
volume density of luminous ($L_{z=3}^*$, or $\sim$$-$21 mag) galaxies
from these searches.  We find that the volume density of luminous
star-forming galaxies at $z\sim7$ is 13$_{-5}^{+8}$$\times$ lower than
at $z\sim4$ and $>$25$\times$ lower ($1\sigma$) at $z\sim9$ than at
$z\sim4$.  This is the most stringent constraint yet available on the
volume density of $\gtrsim L_{z=3}^{*}$ galaxies at $z\sim9$.  The
present wide-area, multi-field search limits cosmic variance to
$\lesssim$20\%.  The evolution we find at the bright end of the $UV$
LF is similar to that found from recent Subaru Suprime-Cam, HAWK-I or
ERS WFC3/IR searches.  The present paper also includes a complete
summary of our final $z\sim7$ $z_{850}$-dropout sample (18 candidates)
identified from all NICMOS observations to date (over the two GOODS
fields, the HUDF, galaxy clusters).
\end{abstract}

\section{Introduction}

The recent WFC3/IR camera on the Hubble Space Telescope (Kimble et
al.\ 2006) has completely revolutionized our ability to search for
galaxies at $z\gtrsim7$ due to its extraordinary imaging capabilities
in the near-infrared -- allowing for large areas to be surveyed to
great depths.  Already some 40 credible $z$$\sim$7-8 galaxy candidates
have been identified in the first hundred orbits of observations
(Oesch et al.\ 2010a; Bouwens et al.\ 2010a,b; McLure et al.\ 2010;
Bunker et al.\ 2010; Yan et al.\ 2010; Finkelstein et al.\ 2010;
Wilkins et al.\ 2010a,b).  This compares with $\sim$15 credible
candidates reported thus far from deep, wide-area ground-based
observations (Ouchi et al.\ 2009; Castellano et al.\ 2010; Hickey et
al.\ 2010) and $\sim$12 thusfar with NICMOS (e.g., Bouwens et
al.\ 2008, 2009a; Bradley et al.\ 2008; Oesch et al.\ 2009; Zheng et
al.\ 2009; $\sim2$ from Richard et al.\ 2008).  Whereas $\sim$100
orbits of NICMOS observations were required to obtain 1 $z\sim7$
credible galaxy candidate (see e.g. \S4 of Bouwens et al.\ 2009a),
only $\sim$2.5 orbits of WFC3/IR observations are required to find a
similar $z\sim7$ candidate.

Despite these significant advances in our observational capabilities
with WFC3/IR to reach deep and identify large numbers of faint
$z\gtrsim7$ galaxies, a full characterization of the galaxy population
at $z\sim7$ requires that we identify large numbers of galaxies at
\textit{both} high and low luminosities.  All but $\sim$6 galaxies in
early selections of $z$$\sim$7-8 galaxies from early WFC3/IR
observations over the ERS/HUDF09 fields have magnitudes faintward of
26.5 mag (e.g., Oesch et al.\ 2010a; Wilkins et al.\ 2010a; Bouwens et
al.\ 2010c).  As such, it is somewhat challenging to characterize the
properties of relatively luminous galaxies at $z\gtrsim6.5$, and some
expansion of the number of sources known brightward of 26.5 mag would
be beneficial.  Such samples are particularly valuable over fields
such as GOODS (Giavalisco et al.\ 2004) where other valuable
multiwavelength data exist like deep IRAC (Dickinson \& GOODS Team
2004) or Chandra coverage (Brandt et al.\ 2001; Rosati et al.\ 2002).

\begin{figure*}
\epsscale{1.15}
\plotone{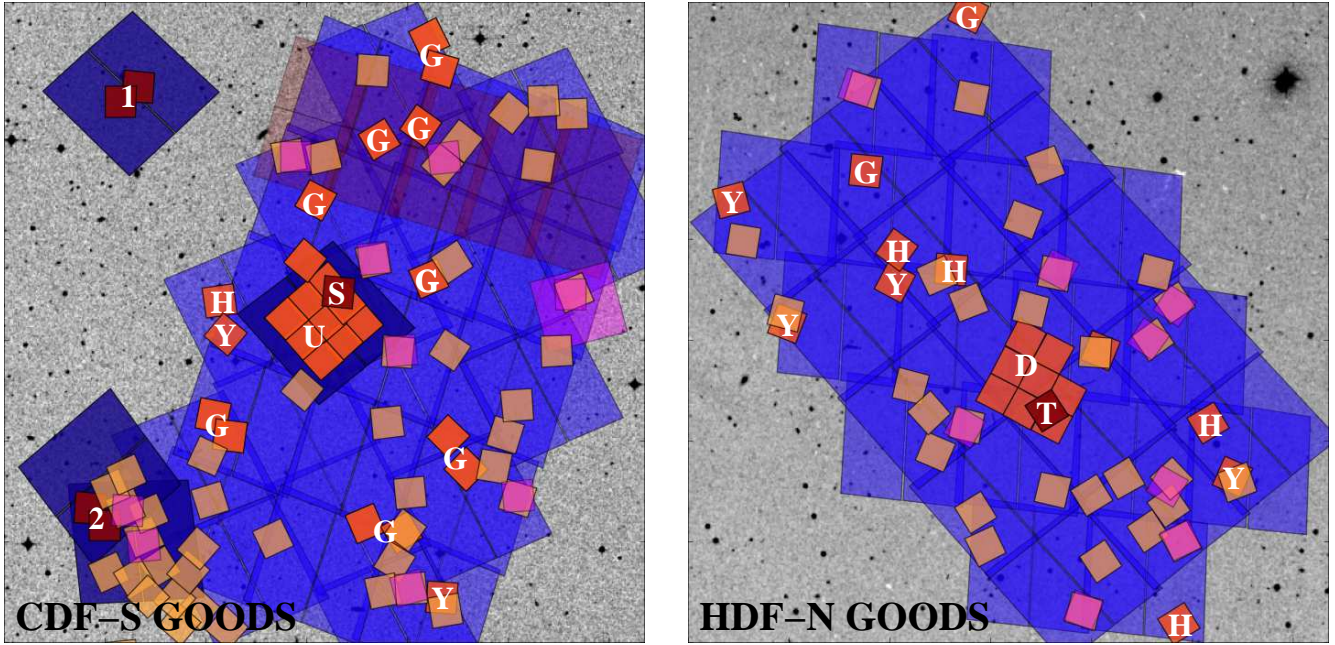}
\caption{\scriptsize{Deep, wide-area near-IR data available to search for
  $z\gtrsim7$ galaxies over the CDF-South GOODS (\textit{left}) and
  HDF-North GOODS (\textit{right}) fields (see also
  Table~\ref{tab:obsdata}).  Most of the ultra-deep NICMOS and deep
  NICMOS observations were already presented in Bouwens et al.\ (2008)
  and included here on this figure (red and dark orange regions
  corresponding to regions with $5\sigma$ depths of $\gtrsim28$ mag
  and $\gtrsim26.5$ mag, respectively).  What is new in the current
  analysis is the moderately deep (26.9 mag at $5\sigma$),
  wide-area ($\sim$58 arcmin$^2$: $>$60 NIC3 pointings) NICMOS
  $H_{160}$-band data (light orange squares: Conselice et
  al.\ 2010).  These data can be used to search for
  $z\gtrsim7$ galaxy candidates by identifying those sources that are
  not detected in the ACS optical bands, but have very red
  $z_{850}-H_{160}$ colors and possess reasonably blue $H-5.8\mu m$ and
  $3.6\mu m-5.8\mu m$ colors.  Unfortunately, these criteria (while very
  demanding) are not sufficient to place strong enough constraints on
  the redshift of the candidates, and so we also obtained deep (27.0
  mag at $5\sigma$) NICMOS $J_{110}$-band imaging (\textit{magenta
    squares}) over the best $z\gtrsim7$ candidates
  (Table~\ref{tab:candlist}: see \S3.2) with GO program 11144 (PI:
  Bouwens).  Also shown in dark orange -- and annotated with ``Y'' or
  ``H'' (for the GO11192 H. Yan et al.\ [2010, in prep] or Henry et
  al.\ 2009 fields, respectively) -- are several additional NICMOS
  fields we used to search for $z\gtrsim7$ galaxies not
  considered by Bouwens et al.\ (2008).  NICMOS fields
  previously considered by Bouwens et al.\ (2008) in searches for
  $z\gtrsim7$ galaxies are indicated as follows: ``D'' denotes the
  HDF-North Dickinson field (Dickinson 1998), ``T'' denotes the
  HDF-North Thompson field (Thompson et al.\ 1999), ``U'' denotes the
  HUDF Thompson field (Thompson et al.\ 2005), ``S'' denotes the
  HUDF05 NICMOS field over the HUDF (Oesch et al.\ 2009), ``1''
  denotes the first set of NICMOS parallels to the HUDF (NICP12: Oesch
  et al.\ 2009), ``2'' denotes the second set of NICMOS parallels to
  the HUDF (NICP34: Oesch et al.\ 2009), and ``G'' denotes the GOODS
  Parallel NICMOS fields.  The blue and dark blue regions correspond
  to regions with deep and very deep optical ACS coverage,
  respectively ($5\sigma$ depths of $\gtrsim28$ and $\gtrsim29$ 
  mag).  The position of the ground-based ISAAC+MOIRCS search areas is
  not shown here to minimize confusion (but is presented in Figure 1
  of Bouwens et al.\ 2008).  Also not included on this figure are the
  NICMOS search fields (Zirm et al.\ 2007) over the HDF South
  (Williams et al.\ 2000).  The position of the Early Release Science
  WFC3/IR observations with the CDF South (not used here for $z\sim7$
  LF constraints) is indicated by the light shaded red region.  A link
  to our NICMOS reductions over the GOODS fields is provided at
  http://firstgalaxies.org//astronomers-area.}
\label{fig:obsdata}}
\end{figure*}

Fortunately, for the selection of luminous $z$$\sim$7-8 galaxies, some
$\sim$88 arcmin$^2$ of deep ($>$26.5 mag, $5\sigma$), wide-area NICMOS
observations exist over the two GOODS fields and the HDF South.  While
$\sim$23 arcmin$^2$ of those observations have already been used to
identify $z$$\sim$7-8 galaxies (Bouwens et al.\ 2008; Oesch et
al.\ 2009), $\gtrsim$60 arcmin$^2$ of those data have yet to be used.
Most of these data are associated with the GOODS NICMOS survey
(Conselice et al.\ 2010) or are NICMOS parallels associated with other
HST programs.  In total, these NICMOS observations cover 2$\times$ as
much area as available in the WFC3/IR observations that made up the
Early Release Science Program (PI O'Connell: GO 11359).  Meanwhile,
these observations cover comparable area to that available in the
wide-area HAWK-I $Y$-band observations over the CDF-South (Castellano
et al.\ 2010; Hickey et al.\ 2010) and less area than the Subaru
Suprime-Cam (Ouchi et al.\ 2009) observations over the HDF-North, but
are deeper on average (by $\sim$0.2 and $\sim$0.7 mag, respectively).

In the present work, we use these wide-area NICMOS observations to
identify a small sample of luminous star-forming galaxies at $z\sim7$,
adding significantly to that known, and performing a search at
$z\sim9$.  Some of the present $z\sim7$ sample has already been used
by Gonzalez et al.\ (2010) and Labb{\'e} et al.\ (2010b) to do stellar
population modelling of luminous $z\sim7$ galaxies and to extend
measures of the stellar mass density and specific star formation rate
to $z\sim7$.  Here we describe the selection of those $z\gtrsim7$
candidates in detail, discuss the properties and layout of the NICMOS
observations, and summarize the properties of the sample.  We also
estimate the contamination rates and the selection volume for this
sample.  Finally, we use this search to derive a constraint on the
volume density of $L_{z=3}^*$ (or $\sim$$-21$ mag) galaxies at
$z\sim7$.  We will also incorporate results from the $\sim$248
arcmin$^2$ Bouwens et al.\ 2008 search for $z\sim7$ galaxies in deep
ground-based data over GOODS.

The structure of this paper is as follows.  In \S2, we summarize the
observational data.  In \S3, we describe our search results for
$z\gtrsim7$ galaxies.  In \S4, we use the observational search results
to derive a constraint on the bright end of the $z\sim7$ and $z\sim9$
$UV$ LFs.  Finally, we provide a summary (\S5).  Throughout this work,
we often denote luminosities in terms of the characteristic luminosity
$L_{z=3}^{*}$ at $z\sim3$ (Steidel et al.\ 1999), i.e.,
$M_{UV,AB}=-21.07$.  Where necessary, we assume $\Omega_0 = 0.3$,
$\Omega_{\Lambda} = 0.7$, $H_0 = 70\,\textrm{km/s/Mpc}$.  Although
these parameters are slightly different from those determined from the
WMAP seven-year results (Komatsu et al.\ 2010), they allow for
convenient comparison with other recent results expressed in a similar
manner.  We express all magnitudes in the AB system (Oke \& Gunn
1983).

\begin{deluxetable*}{ccccccc}
\tablecolumns{7}
\tablecaption{ACS + NICMOS + Ground-Based Imaging data used for our $z$ and $J$ dropout
searches.\tablenotemark{a}\label{tab:obsdata}}
\tablehead{
\colhead{} & \colhead{} & \multicolumn{4}{c}{5$\sigma$ Depth\tablenotemark{b}} & \colhead{} \\
\colhead{Name} & \colhead{Area} & \colhead{$z_{850}$} & \colhead{$J_{110}$} & 
\colhead{$H_{160}$} & \colhead{$K_{s}$} & \colhead{Ref\tablenotemark{c}}}
\startdata
\multicolumn{7}{c}{New NICMOS Fields} \\
GOODS NICMOS Survey & 44.6 & 27.5 & ---\tablenotemark{d} & 26.9 & $\sim$25\tablenotemark{h} & [1] \\
Teplitz Parallels\tablenotemark{e} & 9.3 & 27.5 & ---\tablenotemark{d} & 26.9 & $\sim$25\tablenotemark{h} & [2] \\
HDF-South & 4.3 & 26.4 & 26.3\tablenotemark{f} & 26.7 & 25.7 & [3] \\
Yan Survey & 3.1\tablenotemark{g} & 27.5 & 26.9 & 26.7 & --- & [4] \\
Henry Fields & 3.4\tablenotemark{g} & 27.5 & 26.9 & 26.7 & --- & [5] \\
\multicolumn{7}{c}{NICMOS Fields Already Considered by Bouwens et al.\ (2008)} \\
HDF-North Dickinson & 4.0 & 27.8 & 27.0 & 27.0 & 25.6 & [6,7] \\
HDF-North Thompson & 0.8 & 27.8 & 28.0 & 28.1 & 25.6 & [8,7] \\
HUDF Thompson & 5.8 & 29.0 & 27.6 & 27.4 & 26.0 & [9,10] \\
HUDF Stiavelli & 0.7 & 29.0 & 28.1 & 27.9 & 26.0 & [10,11] \\
HUDF-NICPAR1 & 1.3 & 28.6 & 28.6 & 28.4 & -- & [11,12] \\
HUDF-NICPAR2 & 1.3 & 28.6 & 28.6 & 28.4 & -- & [11,12] \\
GOODS Parallels & 9.3 & 27.5 & 27.0 & 26.9 & $\sim25$\tablenotemark{h} & [12] \\

\multicolumn{7}{c}{Fields with Ground-Based Data Already Considered in
  Bouwens et al.\ (2008)} \\ ISAAC v2.0 (CDFS) & 136 & 27.5 &
$\sim25.4$\tablenotemark{h} & $\sim24.8$\tablenotemark{h} &
$\sim25$\tablenotemark{h} & [13] \\ MOIRCS GTO-2 (HDFN) & 28 & 27.5 &
25.6 & -- & 25.6 & [7] \\ MOIRCS GTO-1,3,4 (HDFN) & 84 & 27.5 & 24.2 &
-- & 24.4 & [7] \\ \enddata \tablenotetext{a}{The layout of these
  search fields is illustrated in Figure~\ref{fig:obsdata}.}
\tablenotetext{b}{$5\sigma$ depths for ACS and NICMOS data given for a
  $0.6''$-diameter aperture and for a $\sim1.0''$-diameter aperture
  for the ground-based $K_s$-band data.  No correction has been made
  for the nominal light outside these apertures (for example, for a
  point source, the correction is typically $\sim$0.2 mag) to keep the
  present estimates as empirical as possible.  This is in contrast to
  the convention that we use in some previous work (e.g., Bouwens et
  al.\ 2008) where such corrections have been made.}
\tablenotetext{c}{References: [1] Conselice et al.\ 2010, 
  [2] Siana et al.\ (2007), [3] Labb{\'e} et al.\ (2003), Zirm et
  al.\ (2007) [4] H. Yan et al.\ (2010, in prep) [5] Henry et
  al.\ (2009), [6] Dickinson 1998, [7] Kajisawa et al.\ 2006, Ouchi et
  al.\ 2007, [8] Thompson et al. 1999, [9] Thompson et al. (2005),
  [10] Labb{\'e} et al.\ (2006), [11] Oesch et al.\ 2007, [12] Bouwens
  \& Illingworth (2006), Riess et al.\ (2007), Siana et al.\ (2007),
     [13] Retzlaff et al.\ 2010, Mannucci et al.\ 2007, Stanway et
     al.\ (2008).}
\tablenotetext{d}{NICMOS $J_{110}$-band observations, with $5\sigma$
    depths of 27.0 mag ($0.6''$-diameter aperature) were acquired
    in those fields with $z\gtrsim7$ candidates.}
\tablenotetext{e}{40-orbit NICMOS $H_{160}$-band observations taken
    in parallel with ACS SBC far-UV observations of the HUDF (Siana et
    al.\ 2007; GO10403: PI Teplitz)}

\tablenotetext{f}{The $J$-band observations here are from
    the deep FIRES observations over the HDF-South with ISAAC
    (Labb{\'e} et al. 2003).}  
\tablenotetext{g}{Not including
    overlap with the GOODS NICMOS Survey (Conselice et al.\ 2010)} 
\tablenotetext{h}{The depth of the near-IR data over the CDF-South
    varies by $\sim$0.2-0.4 mag depending upon the observational
    conditions in which the ISAAC data were taken (Mannucci et
    al.\ 2006; Stanway et al.\ 2008; Retzlaff et al.\ 2010).}
\end{deluxetable*}

\section{Observational Data}

The present search for galaxy candidates at $z\gtrsim7$ makes use of
$\sim$88 arcmin$^2$ of deep NICMOS + ACS observations over the two
GOODS fields + HDF South, as well as $\sim$248 arcmin$^2$ of deep
MOIRCS+ISAAC ground-based data.

Table~\ref{tab:obsdata} provides a convenient summary of the
observational properties of each of these search fields, while
Figure~\ref{fig:obsdata} shows their position within the two GOODS
fields.  A significant part of these observations (e.g., those over
the HDF North [Williams et al. 1996], HUDF [Beckwith et al.\ 2006],
NICMOS parallels to HUDF) have already been used to search for
$z\gtrsim7$ galaxies (e.g., Bouwens et al.\ 2008; Oesch et al.\ 2009),
and so we will not repeat a detailed description of those observations
here.

Instead we focus on the NICMOS observations taken over the past three
years not yet utilized for $z\gtrsim7$ searches.  These observations
include 317 orbits of NICMOS data from five different HST programs
(GO9723, GO10403, GO11082, GO11144, GO11192) over $\gtrsim$60
arcmin$^2$.

The foundation of our search for new $z\gtrsim7$ candidates is
provided by the NICMOS observations of the first three programs
(GO9723, GO10403, and GO11082).  These programs involve 236 orbits of
$H_{160}$-band NICMOS imaging observations, and cover $\sim$58
arcmin$^2$ of the $\sim$65 arcmin$^2$ of new search area.\footnote{We
  note that a small fraction ($\sim$3-4 arcmin$^2$) of this search
  area had been previously considered (Henry et al.\ 2009).}  180 of
these orbits came from the GOODS NICMOS Survey (Conselice et
al.\ 2010), with 60 separate 3-orbit NIC3 pointings.  40 of the orbits
were obtained in parallel with ACS SBC far-UV observations over the
HUDF (Siana et al.\ 2007; GO10403: PI Teplitz), and 16 of the orbits
came from a NICMOS program over the HDF-South WFPC2 field (Labb{\'e}
et al.\ 2003; GO9723: PI Franx), with 8 2-orbit NIC3 pointings.  We
have indicated the position of those NIC3 pointings that lie within
the GOODS fields on Figure~\ref{fig:obsdata} with light orange
squares.  These observations were reduced with our NICMOS pipeline
``nicred.py'' (Magee et al.\ 2007) and reach 26.7 and 26.9 mag
($5\sigma$, 0.6$''$-diameter apertures) depending upon whether the
$H_{160}$-band integrations were 2 or 3 orbits, respectively, in
duration.  The FWHM for the $H_{160}$-band PSF is $\sim$0.37$''$.

\begin{deluxetable*}{cccccccccccc}
\tabletypesize{\tiny}
\tablecolumns{12}
\tablecaption{$z\gtrsim7$ candidates in our wide-area NICMOS $H_{160}$-band data preselected for $J_{110}$-band follow-up observations.\tablenotemark{a}\label{tab:candlist}} 
\tablehead{
  \colhead{} & \colhead{} & \colhead{} & \colhead{NICMOS}
  \\ 
  \colhead{} & \colhead{} & \colhead{} & \colhead{Follow-Up}
  \\ 
\colhead{ID} & \colhead{R.A.} & \colhead{Dec} &
  \colhead{Pointing\tablenotemark{b}} & \colhead{$B-H$} & \colhead{$V-H$} &
  \colhead{$i-H$} & \colhead{$z-H$} & \colhead{$z-J$} &
  \colhead{$J-H$} & \colhead{$H$} & \colhead{$\sigma(H)$\tablenotemark{c}}} 
\startdata
\multicolumn{12}{c}{$z\sim7$ $z_{850}$-dropout candidates} \\ 
GNS-zD1\tablenotemark{d} & 03:32:43.29 & $-$27:42:47.9 & S4 & $>$3.2 & $>$3.7 & 2.7$\pm$0.7 & 1.7$\pm$0.3 &
1.3$\pm$0.4 & 0.4$\pm$0.2 & 25.8$\pm$0.2 & 9 \\ 
GNS-zD2 & 03:32:32.03 & $-$27:45:37.2 & S5 & $>$2.6 & 2.9$\pm$0.6 & $>$2.3 & $>$2.2 & 1.6$\pm$1.0
& 0.6$\pm$0.3 & 26.2$\pm$0.3 & 5 \\ 
GNS-zD3\tablenotemark{e,f} & 03:32:06.10 & $-$27:46:37.3 & S2 & $>$3.1 & $>$3.4 & $>$2.9 & $>$2.5 & 1.5$\pm$1.0 & 0.8$\pm$0.5 &
26.1$\pm$0.3 & 5 \\ 
GNS-zD4 & 12:36:10.93 & 62:09:15.6 & N3 & $>$2.6 & $>$3.0 & $>$2.4 & $>$2.3 & $>$1.6 & 0.7$\pm$0.3 & 25.8$\pm$0.3 & 6
\\ 
GNS-zD5\tablenotemark{f} & 12:36:44.68 & 62:16:15.4 & N4 & $>$3.3 & $>$3.6 & $>$3.2 &
$>$2.5 & 1.7$\pm$0.8 & 1.0$\pm$0.3 & 25.4$\pm$0.2 & 10 \\ 
GNS-zD6\tablenotemark{d,g} & 03:32:22.66 & $-$27:43:00.6 & S1 & $>$3.2 & $>$3.4
& $>$2.6 & 1.9$\pm$0.4 & 1.6$\pm$0.5 & 0.2$\pm$0.2 & 25.4$\pm$0.2
& 11 \\ 
GNS-zD7\tablenotemark{d,h} & 03:32:42.84 & $-$27:42:47.7 & S4 & $>$2.5 & $>$3.0 & $>$2.3 & 1.9$\pm$0.5 & 1.5$\pm$1.0 & 0.4$\pm$0.3 & 26.1$\pm$0.2 & 5
\\ 
\multicolumn{12}{c}{$z\sim9$ $J$-dropout candidates} \\ 
GNS-JD1\tablenotemark{i} & 03:32:13.77 & $-$27:52:42.8 & S8 & --- & 2.9$\pm$0.9 & 2.5$\pm$0.9 &
$>$1.9 & $>$0.5 & 1.4$\pm$0.5 & 25.7$\pm$0.2 & 6 \\ 
GNS-JD2\tablenotemark{j} & 12:36:25.47 & 62:14:31.8 & N2 & $>$3.3 & $>$3.4 & $>$2.5 & $>$2.3 & --- & $>$2.0 & 25.8$\pm$0.3 & 6 \\ 
\multicolumn{12}{c}{Probable Spurious $J$-dropout Candidates\tablenotemark{k}} \\ 
GNS-Sp1 & 03:33:04.66 & $-$27:52:27.0 & S3 & --- & $>$3.4 & $>$3.2 & 2.3$\pm$0.5 & --- & $>$2 & 26.5$\pm$0.3 & 5 \\ 
GNS-Sp2 & 03:33:04.14 & $-$27:52:57.4 & S3 & --- & $>$3.4 & $>$3.0 & $>$2.3 & --- & $>$2 & 26.0$\pm$0.2 & 5 \\ 
GNS-Sp3 & 03:32:08.06 & $-$27:46:58.1 & S2 & $>$2.9 & $>$3.1
& $>$2.9 & $>$2.4 & --- & $>$2.1 & 27.0$\pm$0.3 & 5 \\ 
\multicolumn{12}{c}{Other Candidates Targetted in Preselection\tablenotemark{l}} \\ 
GNS-O1 & 03:33:02.18 & $-$27:53:40.4 & S6 & $>$2.0 & $>$2.2 & 0.8$\pm$0.6 &
1.7$\pm$0.8 & 0.9$\pm$0.8 & 0.7$\pm$0.5 & 26.7$\pm$0.3 & 5 \\ 
GNS-O2 & 03:32:27.81 & $-$27:54:48.0 & S7 & 2.7$\pm$0.6 & $>$3.4 &
$>$2.9 & 1.4$\pm$0.4 & 0.3$\pm$0.5 & 1.1$\pm$0.4 & 25.1$\pm$0.2 & 7
\\ 
GNS-O3 & 12:36:15.21 & 62:10:39.7 & N6 & $>$2.3 & $>$2.5 & $>$2.0 & 1.4$\pm$0.5 & 0.6$\pm$0.7 & 0.8$\pm$0.6 & 26.7$\pm$0.3 & 5 \\ 
GNS-O4 & 12:36:11.26 & 62:09:00.0 & N3 & 2.1$\pm$0.6 & $>$2.8 & $>$2.2 &
1.3$\pm$0.5 & 0.3$\pm$0.6 & 0.9$\pm$0.5 & 26.3$\pm$0.3 & 5 \\ 
GNS-O5 & 12:36:19.14 & 62:15:23.4 & N1 & 2.8$\pm$0.9 & $>$3.0 & $>$2.6 &
1.8$\pm$0.6 & 0.6$\pm$0.7 & 1.2$\pm$0.4 & 25.5$\pm$0.2 & 5 \\ 
GNS-O6 & 12:36:31.68 & 62:06:45.8 & N5 & $>$2.7 & 2.6$\pm$0.6 & $>$2.4 & $>$2.3 & --- & $>$2.0 & 26.2$\pm$0.3 & 5\\ 
GNS-O7 & 12:37:06.51 & 62:11:49.0 & N7 & $>$2.8 & $>$3.0 & $>$2.6 & 2.1$\pm$0.6 & $<$$-$0.2 & $>$1.9 & 26.1$\pm$0.3 & 5 \\ 
GNS-O8 & 12:37:03.01 & 62:11:35.8 & N7 & $>$3.0 & $>$3.1 & $>$2.7 & 1.2$\pm$0.3 & 0.3$\pm$0.4 & 0.9$\pm$0.3 & 26.0$\pm$0.3 & 6 \\ 
\enddata
\tablenotetext{a}{$\,$Lower limits are $1\sigma$.}
\tablenotetext{b}{NIC3 pointing (from GO11144) used for $J_{110}$-band
  follow-up observations on specific $z\gtrsim7$ candidates.
  Fourteen different NIC3 pointings were utilized (S1, S2, S3, S4, S5,
  S6, S7, S8, N1, N3, N4, N5, N6, N7).}

\tablenotetext{c}{Significance (in $\sigma$) of the $H_{160}$-band
  detection.  The significance is measured in our smaller scalable
  apertures (Kron factor of 1.2).  These apertures are smaller than
  those used for estimates of the total magnitudes (\S3.1), and hence
  the associated significance levels are accordingly higher.}

\tablenotetext{d}{Flux information is also available on these
  candidates from the WFC3/IR ERS program (see
  Table~\ref{tab:wfc3comp}).}

\tablenotetext{e}{WFC3/IR $J_{125}$-band data is also available for
  GNS-zD3 as a result of our GO 11144 program.  The measured
  $J_{125}$-band magnitude for this candidate is 26.5$\pm$0.2 mag.
  The $J_{125}-H_{160}$ color (0.4$\pm$0.3) is somewhat red for a
  $z\sim7$ galaxy.  GNS-zD3 is thus more likely at low redshift than
  most candidates in our selection.}
\tablenotetext{f}{These $z\sim7$ candidates are sufficiently red in
  their $J_{110}-H_{160}$ colors that there is an increased
  probability they could be low redshift contaminants.  We have no
  evidence for this possibility however.\\
 $^g$ Also identified in Castellano et al.\ (2010) and
  Hickey et al.\ (2010)\\
 $^h$ This source has a measured $z_{850}-Y_{105}$ color
  $\sim0.0\pm1.0$ and $J_{125}-H_{160}\sim0.6\pm0.2$ color in the high
  S/N WFC3/IR observations (see Table~\ref{tab:wfc3comp}).  This
  suggests this candidate may be a red $z$$\sim$1-2 galaxy
  and not a $z\sim7$ star-forming galaxy.\\
 $^i$ GNS-JD1 is detected at $1\sigma$ in the $V_{606}$
  and $i_{775}$ bands, not sufficient to rule it out as a $z\sim9$
  $J_{110}$ dropout candidate, but suggesting it may lie at
  $z$$\sim$1-2.\\
 $^j$ We have no evidence that GNS-JD2 is detected at
  wavelengths other than 1.6$\mu m$ ($H$-band).  It is sufficiently
  close to another source that it is unclear if it is detected redward
  of 2$\mu m$ from the IRAC data.  This may suggest it is a transient
  source (SNe: see \S3.4) or spurious (since it is close to the edge
  of the NIC3 field where it was found).  However, we cannot rule out
  the possibility it corresponds to a $z\sim9$ galaxy (but we consider
  it unlikely).\\
 $^k$ These candidates were targetted for follow-up
  observations based upon $\sim$5$\sigma$ detections in the NICMOS
  $H_{160}$-band data.  However, they are not detected in the NICMOS
  $J_{110}$ band observations.  Lacking such detections, their reality
  is less secure.  We suspect that they may be spurious (see \S3.4).\\
 $^l$ These candidates were targetted for $J_{110}$-band
  follow-up observations with NICMOS (because of their red
  $z_{850}-H_{160}>1.2$ colors, apparent absence of flux in the
  optical, and moderately blue $H-5.8\mu m$ colors: see \S3.1).
  However, they do not satisfy our $z\sim7$ $z_{850}$-dropout or
  $z\sim9$ $J_{110}$-dropout criteria.}
\end{deluxetable*}

The first two of these programs (GO10403 and GO11082) had deep optical
coverage from the GOODS ACS program, with corresponding depths
reaching 27.9, 28.1, 27.4, and 27.3 in the $B_{435}$, $V_{606}$,
$i_{775}$, and $z_{850}$ bands, respectively, in 0.6$''$-diameter
apertures.  Our reductions of the ACS GOODS observations are described
in Bouwens et al.\ (2007) and take advantage of essentially all ACS
data ever taken over the GOODS fields.  They are similar to the GOODS
v2.0 reductions, but reach $\sim$0.1-0.3 mag deeper in the
$z_{850}$-band due to the inclusion of the SNe follow-up data (Riess
et al.\ 2007).  The third program here (GO9723) had deep WFPC2
observations from the HDF South program (Williams et al.\ 2000), and 2
orbit $i_{775}$/2 orbit $z_{850}$ ACS WFC observations as part of the
ACS GTO program (PI Ford: GTO 9301).

The IRAC reductions we use for our selection of $z\gtrsim7$ galaxy
candidates are those from Dickinson \& GOODS Team (2004).  They reach
to 27.4 mag, 26.8 mag, 25.4 mag, and 25.3 mag in the $3.6\mu m$,
$4.5\mu m$, $5.8\mu m$, and $8.0\mu m$ bands, respectively, at $1\sigma$
($2''$) in most regions (where 23 hour integrations are considered)
and 0.4 mag deeper (where the integration time is 46 hours).

Deep $J_{110}$-band observations became available over some fraction
of these fields, as a result of our 60-orbit GO11144 NICMOS program.
The $J_{110}$-band observations were $\sim$2-3 orbits in duration,
reaching depths of 26.9-27.0 mag ($5\sigma$ in $0.6''$-diameter
aperture) and targetted specific $z\gtrsim7$ candidates identified in
the original ACS optical + NICMOS $H_{160}$ band observations (see \S
3.2 for a description of the preselection).  Only 14 such NIC3 fields
were obtained and are indicated in Figure~\ref{fig:obsdata} with the
magenta squares (though two of the NIC3 fields shown targetted
specific $z$$\sim$6 candidates to allow for a measurement of the
$UV$-continuum slope $\beta$: see Bouwens et al.\ 2009b).  Again the
reduction of those observations was performed with ``nicred.py''
(Magee et al.\ 2007), an alignment made to the ACS GOODS data, and
then the data drizzled onto the ACS GOODS frame rebinned on a
0.09$''$$\times$0.09$''$ frame.  The FWHM for the $J_{110}$-band PSF
is $\sim$0.33$''$.

Two of our NIC3 fields chosen for $J_{110}$-band follow-up were
allocated more than 3 orbits of time, to better ascertain the nature
of the $z\gtrsim7$ candidates in those fields.  In the first case (S4:
03:32:42.48, $-$27:42:48.8), additional follow-up observations were
obtained with NICMOS.  In the other case (S2: 03:32:06.75,
$-$27:47:07.0), additional follow-up observations were obtained with
WFC3/IR in the $J_{125}$ band (1 orbit).

While most of the search power of our program is provided by the first
three $H_{160}$-band programs and the follow-up $J_{110}$-band
program, we also include observations from two other HST programs in
our search for $z\gtrsim7$ galaxies.  One of these programs is the
H. Yan et al. (2010, in prep) GO 11192 program designed to follow up
on bright $z_{850}$-dropout candidates identified in wide-area
ground-based near-IR observations over the two GOODS fields (see the
dark orange squares in Figure~\ref{fig:obsdata} annotated with ``Y'').
In that program, 4 orbit $J_{110}+H_{160}$ band observations (2 orbits
in the $J_{110}$ band and 2 orbits in the $H_{160}$ band) were
obtained on 6 different bright $z\sim7$ candidates.  These fields
reached depths of 26.9 and 26.7 mag in the $J_{110}$ and $H_{160}$
bands, respectively, in 0.6$''$-diameter apertures.

The other program that we utilized was GO10872 (Henry et al.\ 2009;
Siana et al.\ 2010).  NICMOS parallel observations from that program
cover $\sim$3.4 arcmin$^2$ of area and lie within the two GOODS fields
(see the dark orange squares in Figure~\ref{fig:obsdata} annotated
with ``H'').  These observations reach to $\sim$26.9 mag in the
$J_{110}$ band and $\sim$26.7 mag in the $H_{160}$ band.  While Henry
et al.\ (2009) have already used these fields to search for candidate
$z\gtrsim7$ galaxies -- reporting none -- we shall nevertheless
incorporate these observations into our $z\gtrsim7$ search to be as
comprehensive as possible.

\section{Selection of $z\gtrsim7$ Candidates}

Here we describe the selection of $z\gtrsim7$ galaxies over those
NICMOS fields not previously considered by our team for such searches
($\sim$65 arcmin$^2$ in search area).  We will not revisit the
searches for $z\gtrsim7$ galaxies conducted by Bouwens et al.\ (2008)
over the $\sim$23 arcmin$^2$ of area within the HUDF, HDF North, and
CDF South or the $\sim$248 arcmin$^2$ of deep ground-based data.
Instead we will simply incorporate the Bouwens et al.\ (2008) search
results in with our new search results.

\subsection{Catalogues}

We use the same procedure for doing object detection and photometry as
we have used in previous work (e.g., Bouwens et al.\ 2007; Bouwens et
al.\ 2008).  Briefly, we begin by PSF-matching all of our data to the
PSF of our detection image (which here is the NICMOS $H_{160}$ band).
SExtractor (Bertin \& Arnouts 1996) is then run in double-image mode,
with detection image taken to be the square root of the $\chi^2$ image
(Szalay et al.\ 1999).  Here the square root of the $\chi^2$ image is
simply the NICMOS $H_{160}$-band imaging data.  Colors were measured
in a smaller scalable aperture using Kron-style (1980) photometry,
with a Kron factor of 1.2 to improve the S/N (typical aperture
diameters for color measurements were $0.4''$).  To correct the small
aperture flux measurements to total magnitudes, we computed
corrections by comparing the light in a larger scalable (Kron factor
of 2.5) aperture on the $\chi^2$ image with that in the smaller
aperture (Kron factor of 1.2).  Figure 5 of Coe et al.\ (2006)
provides a graphical description of a similar multi-stage procedure
for measuring colors and total magnitudes.  We also make a small
correction for light outside the larger scalable apertures and on the
wings of the PSF (typically $\sim$0.15 mag).  This correction is made
based upon the encircled energy expected to lie outside this aperture
(for stars).

\begin{figure}
\epsscale{1.15}
\plotone{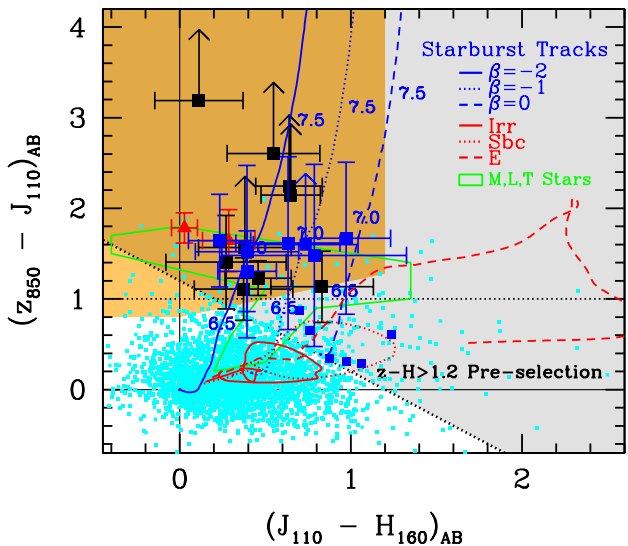}
\caption{\small{$z_{850}-J_{110}$ vs. $J_{110}-H_{160}$ color-color
    diagram used to select $z\sim7$ $z_{850}$-dropouts.  The orange
    region shows the location of $z\sim7$ candidates in color-color
    space.  The expected colors of high-redshift star-forming galaxies
    and low-redshift galaxy contaminants (shown over the redshift
    range $z\sim0$-2) are indicated with blue and red lines, as a
    function of redshift.  The plotted uncertainties and limits are
    $1\sigma$ (consistent with previous work).  The green line enclose
    the region in color-color space where we expect low-mass stars to
    be found (e.g., Knapp et al.\ 2004) while the red triangles show
    the position of two probable T dwarfs identified within the GOODS
    fields (Bouwens et al.\ 2008; Mannucci et al.\ 2006).  The cyan
    points show the colors of individual sources in our search fields.
    The lightly-shaded gray region shows the $z_{850}-H_{160}>1.2$
    preselection we use to identify possible $z\gtrsim7$ candidates to
    follow up with deeper $J_{110}$-band NICMOS observations (\S3.2).
    It is apparent that this preselection identifies all eight
    $z\sim7$ candidates (\textit{large black squares}) previously
    found by Bouwens et al.\ (2008) and the Bradley et al.\ (2008)
    $z\sim7.6$ galaxy.  The large blue squares show the colors of new
    sources that satisfied our $z\sim7$ $z_{850}$-dropout criteria and
    are included in our $z\sim7$ $z_{850}$-dropout sample
    (Tables~\ref{tab:candlist} and ~\ref{tab:zcandlist}) -- while the
    small blue squares show the colors of preselected galaxies that
    did not (and hence are more likely at low redshift).  See
    Table~\ref{tab:candlist} for our photometry on those $z\gtrsim7$
    candidates identified for $J_{110}$-band follow-up
    observations.}\label{fig:zjjh}}
\end{figure}

\subsection{$z\gtrsim7$ Candidate Preselection}

Searches for star-forming galaxies at high-redshift tend to be
relatively straightforward in execution.  Typically these searches
involve the acquisition of imaging data in three adjacent bands
followed by a traditional two-color Lyman-Break Galaxy (LBG)
selection.  While useful, this approach is limited by the available
resources.  Often times, much larger areas can be searched by
utilizing suboptimal data sets that still allow for the efficient
identification of high-redshift star-forming galaxies.

With the availability of deep wide-area NICMOS $H_{160}$-band data
over the two GOODS + HDF-South fields, we have such a data set.  The
wide-area NICMOS + optical + IRAC data allow us to identify sources
that have a high probability of corresponding to $z\gtrsim7$ galaxies.
We can perform such a preselection due to the unique colors of
star-forming galaxies at $z\gtrsim7$.  $z\gtrsim7$ galaxies are very
red in the optical to $H_{160}$ colors and moderately blue in the
$H_{160}$ to IRAC colors.

We then follow up these preselected candidates with deep imaging in
the $J_{110}$-band to better ascertain their nature using more
traditional two-color LBG criteria.  For $z\sim7$ galaxies (\S3.3), we
apply the same $z_{850}-J_{110}$ / $J_{110}-H_{160}$ criterion
utilized by Bouwens et al.\ (2004), Bouwens \& Illingworth (2006), and
Bouwens et al.\ (2008).  Candidates with very red $z_{850}-J_{110}$
colors (corresponding to the Ly$\alpha$ cut-off) and blue
$J_{110}-H_{160}$ colors (corresponding to the UV continuum) are taken
to be $z\sim7$ galaxies.

For the $z\gtrsim7$ preselection, we employ Lyman-break-like criteria.
The candidates must be detected at $\geq5\sigma$ in the $H_{160}$
band, have $(z_{850}-H_{160})_{AB}$ colors redder than 1.2, be
completely undetected ($<2\sigma$) in the ACS $B_{435}$, $V_{606}$,
and $i_{775}$ bands, and not have $H_{160}-5.8\mu m$ colors redder
than 2.5 mag (to exclude dust reddened galaxies at $z$$\sim$1-2).  Our
use of a $z_{850}-H_{160}>1.2$ preselection was designed to ensure
that the candidates we identified showed a prominent break across the
$z_{850}$ and $H_{160}$ bands (as expected for $z\gtrsim7$ dropouts).
However, this criterion was not so strong as to exclude any $z\sim7$
$z_{850}$-dropout galaxies identified in previous NICMOS searches
(e.g., Bouwens et al.\ 2008; Oesch et al.\ 2009) or to have a
substantial impact on the effective selection volume of our
$z\gtrsim7$ search.

Sources that were detected at $1.5\sigma$ in more than two optical
bands ($B_{435}$, $V_{606}$, or $i_{775}$) were also excluded.
Figure~\ref{fig:zjjh} illustrates this preselection in terms of the
$z_{850}-J_{110}$/$J_{110}-H_{160}$ standard two-color diagram used to
identify $z\sim7$ $z_{850}$-dropout galaxies.

\begin{figure}
\epsscale{1.15}
\plotone{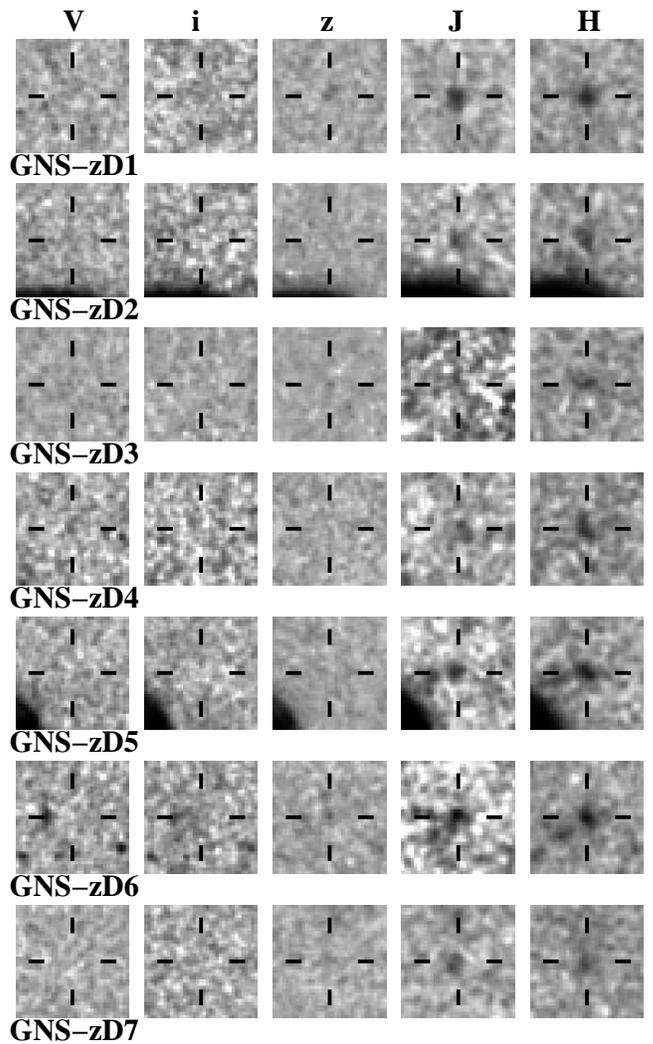}
\caption{$V_{606}i_{775}z_{850}J_{110}H_{160}$ images
  ($3.5''\times3.5''$) of the 7 $z\sim7$ $z_{850}$-dropout candidates
  identified in our new wide-area NICMOS data.  See
  Table~\ref{tab:candlist} for the coordinates, magnitudes, colors,
  and other properties of the present $z_{850}$-dropout candidates.
  Deeper $J_{125}$-band observations with the WFC3/IR camera were
  obtained over GNS-zD3 to confirm the weak $J_{110}$-band detection
  obtained with NICMOS.  The nature of GNS-zD7 is unclear.  While the
  source satisfies our $z_{850}$-dropout criterion (and would
  therefore appear to be a plausible $z\sim7$ galaxy), its $Y-J$,
  $J-H$ colors measured with WFC3/IR suggest a red $z\sim1-2$
  galaxy.\label{fig:stampwide}}
\end{figure}

\begin{deluxetable*}{cccccc|ccc}
\tablecolumns{9} 
\tabletypesize{\scriptsize}
\tablecaption{Available WFC3/IR photometry for $z\sim7$ $z_{850}$-dropout galaxy
  candidates in our wide-area NICMOS sample\label{tab:wfc3comp}} 
\tablehead{
  \colhead{} & \colhead{} & \colhead{} & \colhead{} & \multicolumn{2}{c}{NICMOS} & \multicolumn{3}{c}{WFC3/IR} \\
  \colhead{Object ID} & \colhead{R.A.} & \colhead{Dec} & \colhead{$H_{160}$} & 
  \colhead{$z_{850}-J_{110}$} & \colhead{$J_{110}-H_{160}$} & 
  \colhead{$z_{850}-Y_{098}$} & \colhead{$Y_{098}-J_{125}$} & 
  \colhead{$J_{125}-H_{160}$}}
\startdata
GNS-zD1 & 03:32:43.29 & $-$27:42:47.9 & 25.8$\pm$0.2 & 1.3$\pm$0.4 & 0.4$\pm$0.2 & 0.8$\pm$0.4 & 0.6$\pm$0.3 & 0.2$\pm$0.1 \\
GNS-zD6 & 03:32:22.66 & $-$27:43:00.6 & 25.4$\pm$0.2 & 1.6$\pm$0.5 & 0.2$\pm$0.2 & 1.4$\pm$0.4 & 0.0$\pm$0.2 & 0.4$\pm$0.1 \\
GNS-zD7\tablenotemark{a} & 03:32:42.84 & $-$27:42:47.7 & 26.1$\pm$0.3 & 1.5$\pm$1.0 & 0.4$\pm$0.3 & 0.5$\pm$0.9 & 0.5$\pm$0.6 & 0.7$\pm$0.3 \\
\enddata
\tablenotetext{a}{The measured $z_{850}-Y_{105}$ color $\sim0.0\pm1.0$
  and $J_{125}-H_{160}\sim0.6\pm0.2$ color suggests this candidate may
  be a red $z$$\sim$1-2 galaxy and not a $z\sim7$ star-forming galaxy.}
\end{deluxetable*}

We identified 20 $z\gtrsim7$ candidates for follow-up study from the
wide-area ($\sim$65 arcmin$^2$) NICMOS $H_{160}$-band observations.
The coordinates and photometry for candidates are listed in
Table~\ref{tab:candlist}.  After some experimentation, we discovered
we could fit those $z\gtrsim7$ candidates within 14 $52''\times52''$
NIC3 fields (i.e., S1, S2, S3, S4, S5, S6, S7, S8, N1, N3, N4, N5, N6,
N7).

We obtained follow-up $J_{110}$-band NICMOS observations on these
candidates (typically 3 orbits for each NIC3 pointing) with the
60-orbit HST GO program 11144.  Those observations were obtained from
December 19, 2007 to September 9, 2008, with one final orbit of
observations taken on October 26, 2009 with the WFC3/IR camera.  The
position of these follow-up data is illustrated in
Figure~\ref{fig:obsdata} with the magenta squares (and rectangle for
the 1-orbit WFC3/IR observation).

Once the NICMOS $J_{110}$-band observations were obtained, we redid
our photometry on each $z\gtrsim$7 candidate.  This photometry is
included in Table~\ref{tab:candlist} for all 20 $z\gtrsim7$ candidates
identified for follow-up study.  We then applied the $z_{850}$ and
$J_{110}$-dropout selection criteria we describe in the next two
subsections to identify probable star-forming galaxies at $z\sim7$ and
$z\sim9$, respectively, from this preselection.

\subsection{$z\sim7$ $z_{850}$-dropout Selection}

We use the same selection criterion for identifying $z\sim7$
$z_{850}$-dropouts as we did in our previous work on the HUDF and two
GOODS fields (Bouwens et al.\ 2008).  Specifically, we require our
$z_{850}$-dropout candidates to satisfy the criteria
$((z_{850}-J_{110})_{AB})>0.8)\wedge
((z_{850}-J_{110})_{AB}>0.8+0.4(J_{110}-H_{160})_{AB})$ where $\wedge$
represents the logical \textbf{AND} symbol.  In cases of a
non-detection in the dropout band, the flux in the dropout band is set
to its $1\sigma$ upper limit.  This two-color selection is illustrated
in Figure~\ref{fig:zjjh} with the position of the $z\gtrsim7$
candidates from Table~\ref{tab:candlist} included as the large blue
squares.

$z_{850}$-dropout candidates are required to be detected at $5\sigma$
in the $H_{160}$ band (0.6$''$-diameter aperture) to ensure that they
correspond to real sources.  In addition, included in the present
search are also those sources from the NICMOS observations from the
H. Yan et al.\ (2010, in prep) GO 11192 and Henry et al.\ (2009) GO
10872 programs.

In total, 7 sources from our new search fields satisfied our
$z_{850}$-dropout criteria.  All of these candidates were found over
the GOODS NICMOS Survey (Conselice et al.\ 2010), and none from the
Yan et al.\ (2010, in prep) or Henry et al.\ (2009) fields.  Postage
stamps of these candidates are shown in Figure~\ref{fig:stampwide}.
Flux measurements for the candidates are given in
Table~\ref{tab:fluxdata} in Appendix A.  The sources range in
magnitude from $H_{160,AB}\sim25.4$ to 26.1 mag, with most of the
sources being found at $\sim$26 mag.  The surface density of
$z_{850}$-dropout candidates in our fields ($\sim$65 arcmin$^2$)
brightward of 26.5 mag is $\sim$0.1 source arcmin$^{-2}$, very similar
to that found by Bouwens et al.\ (2008).

Overall, the properties ($J_{110}-H_{160}$ color and sizes) of most of
our candidates seem consistent with their being high-redshift
galaxies.  The mean half-light radii (including the effect of the PSF)
is $\sim$0.25$''$, which is similar to the candidates found over the
HUDF (e.g., Bouwens et al.\ 2008).  Meanwhile, the $J_{110}-H_{160}$
color for the sources is 0.6$\pm$0.1 mag.  This is consistent, albeit
somewhat higher, than the $J_{110}-H_{160}\sim0.48$ color expected
from our simulations (Bouwens et al.\ 2008), assuming a $\beta\sim-2$
(Bouwens et al.\ 2010a; Stanway et al.\ 2005; Finkelstein et
al.\ 2010; Bunker et al.\ 2010).  Nonetheless, a few of the candidates
(GNS-zD3, GNS-zD5) in our selection have sufficiently red
$J_{110}-H_{160}$ colors ($\sim$0.8 to $1.0$ mag) that they might be
low-redshift interlopers.

The NICMOS $J_{110}$-band observations available on GNS-zD3 were only
moderately deep ($\sim$2 orbits) and so GNS-zD3 only shows a weak
detection in that band.  Fortunately, we were able to obtain deeper
$J_{125}$-band observations on GNS-zD3 with the WFC3/IR instrument
(\S2).  The new data confirms there is a detection in the $J$ band,
and the measured magnitude 26.5$\pm$0.2 is consistent with what is
found with NICMOS.

Three of our new $z_{850}$-dropout candidates (GNS-zD1, GNS-zD6,
GNS-zD7) are found over the upper region in the CDF-South where deep
wide-area WFC3/IR observations were recently taken as part of the
Early Release Science program (GO11359: PI O'Connell).  Given that
these observations reach $\sim$0.7-1.0 mag deeper than the NICMOS
observations utilized in this study and extend over three bands
$Y_{098}$, $J_{125}$, and $H_{160}$, they are useful for
characterizing the typical dropout candidates found in this search.

What do these deeper data suggest about the candidates in our
selection?  Photometry on the three aforementioned $z_{850}$-dropout
candidates was performed using the new WFC3/IR observations (utilizing
specifically the Bouwens et al.\ (2010c) reductions).  The results are
summarized in Table~\ref{tab:wfc3comp}, and the conclusions are mixed.
GNS-zD6 is clearly a $z\sim6$ galaxy -- though from the measured
colors its redshift is likely in the range $z$$\sim$6.2-6.5.  The
nature of GNS-zD1 is slightly less clear from the data.  Its measured
$z-J$, $Y-J$ colors are consistent with its being either a $z\sim6.5$
galaxy or a red $z$$\sim$1-2 galaxy.  The blue $J-H\sim0.1$ colors of
GNS-zD1 seem to slightly favor the case that it is a $z\gtrsim5$
galaxy.  Finally, GNS-zD7 seems most consistent with being an
intrinsically red $z$$\sim$1-2 galaxy, having very red $J-H\sim0.6$,
$Y-J\sim0.7$ colors.

\begin{figure}
\epsscale{1.15}
\plotone{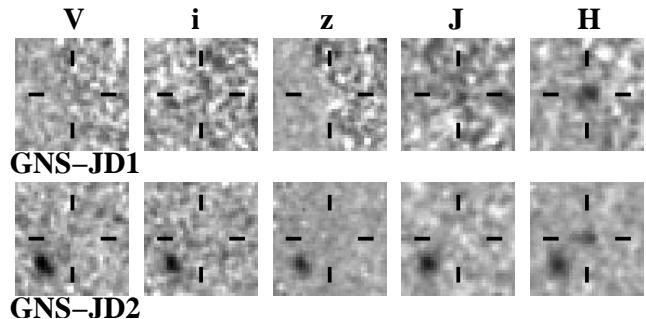}
\caption{$V_{606}i_{775}z_{850}J_{110}H_{160}$ images
  ($3.5''\times3.5''$) of the 2 sources in our new wide-area NICMOS
  data that satisfy our $z\sim9$ $J_{110}$-dropout criteria.  However,
  we consider it unlikely that either of the candidates identified
  here corresponds to a $z$$\sim$9 galaxy.  GNS-JD1 appears to be
  detected at $1\sigma$ in both the $V_{606}$ and $i_{775}$ bands, not
  sufficient for us to rule it out as a $z\sim9$ $J_{110}$-dropout
  candidate but suggesting it may be a $z$$\sim$1-2 galaxy.  GNS-JD2,
  by contrast, shows no evidence for being detected at wavelengths
  other than 1.6$\mu m$ ($H$-band).  This may suggest it is a
  transient source (SNe: see \S3.4) or spurious (since it is close to
  the edge of the NIC3 field where it was found).  However, we cannot
  rule out the possibility it corresponds to a $z\sim9$ galaxy (but we
  consider it very unlikely).  See Tables~\ref{tab:candlist} and
  \ref{tab:fluxdata} for the coordinates, magnitudes, colors, and
  other properties of these candidates.\label{fig:stampwidej}}
\end{figure}

These results suggest that the present $z_{850}$-dropout selection is
successful in identifying $z\gtrsim6.5$ galaxies, albeit with a mean
redshift somewhat lower than the $z\sim7.3$ estimated in Bouwens et
al.\ (2008).  The lower mean redshift for the sample is consistent
with the expected bias based upon evolution across the
$z_{850}$-dropout selection window (where more luminous galaxies are
present e.g. at $z\sim6.5$ than at $z\sim8$: Mu{\~n}oz \& Loeb 2008).
The likely contamination of our selection by one probable low-redshift
source (GNS-zD7) is consistent with the 24\% contamination rates
estimated in \S3.6.

The principal reason we are finding modest levels of the contamination
over the GOODS fields is because of the limited depth of the available
ACS optical data over the GOODS fields.  This contamination is
somewhat higher than estimated over other $z\sim7$ $z_{850}$-dropout
selections like the HUDF -- where it was estimated to be $\sim$12\%
(e.g., Bouwens et al.\ 2008).  In order to reduce these contamination
levels, it would therefore be ideal if deeper optical data --
particularly in the F606W, F775W, and F814W bands -- could be obtained
over the GOODS fields.

\subsection{$z\sim9$ $J_{110}$-Dropout Selection}

Similar to our $z\sim7$ $z_{850}$-dropout selection, we adopt the same
$z\sim9$ $J_{110}$-dropout selection criteria as we used in the
$\sim$23 arcmin$^2$ Bouwens et al.\ (2008) NICMOS search.
$J_{110}$-dropout candidates in our selection are required to satisfy
the criterion $(J_{110}-H_{160})_{AB}>1.3$ and not show $>2\sigma$
detections in any of the optical bands (or $>1.5\sigma$ in two bands).
$z\sim9$ $J_{110}$-dropout candidates are also required to be detected
at 6$\sigma$ in the $H_{160}$ band (0.6$''$-diameter aperture) to
ensure that most of the sources are real.  We use a 6$\sigma$
detection criterion for our $J_{110}$-dropout selection (instead of a
5$\sigma$ criterion) because we only have one passband to evaluate the
reality of the candidates.

In total, we identified 2 sources that satisfied our $z\sim9$
$J_{110}$-dropout criteria.  Postage stamps of these candidates are
provided in Figure~\ref{fig:stampwidej}, and their photometry is
summarized in Table~\ref{tab:candlist}.  Flux measurements for the
candidates are given in Table~\ref{tab:fluxdata} in Appendix A.

While both candidates formally satisfy our $J_{110}$-dropout selection
criteria and therefore may correspond to $z$$\sim$9 galaxies, neither
candidate is very compelling, and there are reasons to suspect each
candidate may be a contaminant.  For example, GNS-JD1 is formally
detected at $1\sigma$ in both the $V_{606}$ and $i_{775}$ bands --
suggesting that it may actually correspond to a $z\sim2$ Balmer-break
galaxy.  In addition, GNS-JD1 is detected at $5\sigma$ in both the
$3.6\mu m$ and 4.5$\mu m$ bands, with $m_{3.6\mu m}=25.1\pm0.2$ mag and
$m_{4.5\mu m}=25.0\pm0.2$ mag.  This latter IRAC photometry was
performed by modelling the light profiles of nearby neighbors in the
IRAC imaging data, subtracting these model profiles from the
observations, and then measuring the flux in simple $2.5''$-diameter
apertures (e.g., Labb\'{e} et al.\ 2006; Gonzalez et al.\ 2010).

The other $J_{110}$-dropout candidate GNS-JD2 is possibly spurious.
None of the observations at other wavelengths provide any evidence it
is real.  It is not detected in the ACS observations, the NICMOS
$J_{110}$-band observations, nor even the IRAC $3.6\mu$ and $4.5\mu$
observations.  One would have expected GNS-JD2 to be formally detected
in the IRAC $3.6\mu$ data at $\sim$1.5-2$\sigma$ if it was well
separated from its neighbors.  However, since the source is not well
separated, flux measurements are sufficiently challenging that the
IRAC non-detection does not definitely argue against the reality of
the source.  Nonetheless, the non-detection does suggest that the
original detection of GNS-JD2 in the $H_{160}$-band may have been
anomalous and therefore the source is likely either spurious or a
transient source (i.e., a SN).

Additional evidence for the source being spurious comes from its
proximity to the edge of the NIC3 field where it was found (due to
enhanced non-Gaussianity of the noise there).  To test the spurious
hypothesis, we examined the individual NICMOS exposures that went into
its $H_{160}$-band stack.  None of these exposures provided a dominant
contribution to the cumulative $H_{160}$-band flux, leaving us with no
compelling reason to flag this particular candidate as spurious.  An
alternate hypothesis is that GNS-JD2 is a SN.  As we discuss in \S3.6,
we would expect $\sim$1 SN to be identified over our $\sim$65
arcmin$^2$ $H_{160}$-band observations as a $z\gtrsim7$ galaxy
candidate.

Besides these 2 formal $J_{110}$-dropout candidates, there were 3
other $z\gtrsim7$ candidates identified in the NICMOS $H_{160}$-band
data for follow-up (GNS-Sp1, GNS-Sp2, GNS-Sp3 in
Table~\ref{tab:candlist}), but which were not detected in the later
$J_{110}$-band observations.  This would give these sources nominal
$J_{110}-H_{160}$ colors $\gtrsim2$ mag ($1\sigma$) and make them
possible $z\sim9$ $J_{110}$-dropout candidates.  However, there are
reasons to be concerned about these candidates and whether they
correspond to real sources.  While the formal significance of each
source is $\sim$5$\sigma$ in the $H_{160}$-band data, these sources
are detected in stacks of $\leq$9 NICMOS exposures (and hence subject
to noise with significant non-Gaussian signatures) and are typically
near a bright source (GNS-Sp1, GNS-Sp3).  Consequently, the true
significance of these sources is somewhat smaller, i.e.,
$\lesssim$4$\sigma$, making it more likely they are spurious (see also
the simulation results in \S3.6).\footnote{Note that this is in
  contrast to situations where similar significance sources are found
  in observations created from a much larger number of exposures
  (e.g., the 56 exposures stacks used for the WFC3/IR HUDF09
  $H_{160}$-band observations: Oesch et al.\ 2010a; Bouwens et
  al.\ 2010b).  In those cases, the noise characteristics are much
  closer to Gaussian, and apparent $5\sigma$ sources are indeed
  significant at the $5\sigma$ level.} These three sources are
included in Table~\ref{tab:candlist} under the label ``Probable
Spurious  $J$-dropout Candidates.''

\begin{figure}
\epsscale{1.15}
\plotone{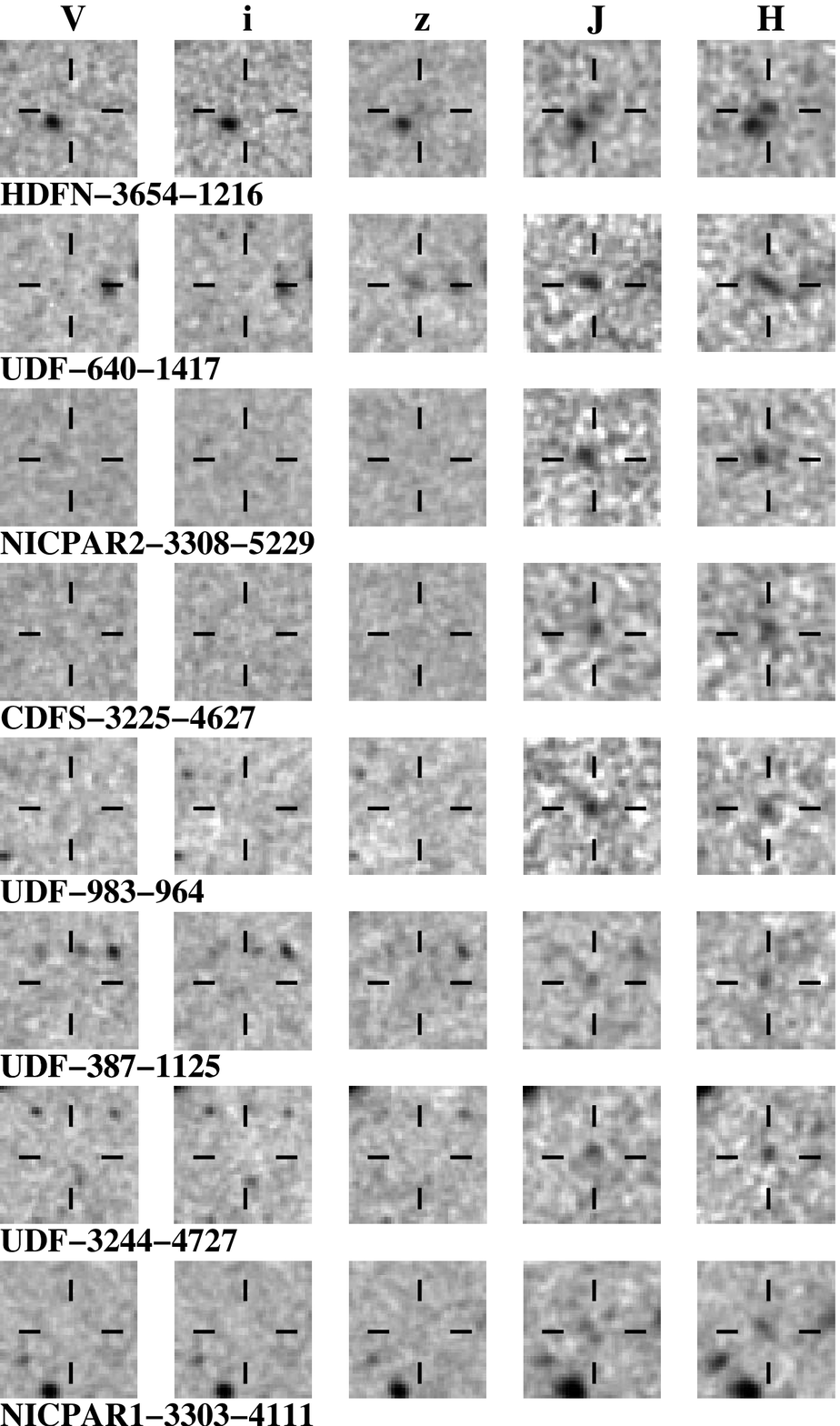}
\caption{$V_{606}i_{775}z_{850}J_{110}H_{160}$ images of the 8
  $z\sim7$ $z_{850}$-dropout candidates previously identified in the
  ultra-deep, wide-area NICMOS data (Bouwens et al.\ 2008: but see
  also e.g. Oesch et al.\ 2009), but now utilizing the deeper
  $J_{110}$-band observations (8 additional orbits) obtained on two
  candidates (UDF-3244-4727 and UDF-387-1125) as part of the GO 11144
  program.  The $J_{110}$-band detections for these two candidates are
  now much more significant than they were in Bouwens et al. (2008:
  Figure 3 from that work).  The other 6 candidates are included here
  for completeness.  Each of the above candidates is detected at
  $\geq4.5\sigma$ in both the $J_{110}$ and $H_{160}$
  bands.  \label{fig:stampdeep}}
\end{figure}

\begin{deluxetable*}{ccccccccc}
\tabletypesize{\scriptsize}
\tablecolumns{9} \tablecaption{Our total sample of $z\sim7$ 
  $z_{850}$-dropout galaxy
  candidates selected from NICMOS observations over the CDF-S + HDF-N GOODS fields, the HUDF, and galaxy cluster fields (see Table~\ref{tab:obsdata}).\tablenotemark{a}\label{tab:zcandlist}}

\tablehead{
  \colhead{Object ID} & \colhead{R.A.} & \colhead{Dec} &
  \colhead{$H_{160}$} & \colhead{$z_{850}-J_{110}$} &
  \colhead{$J_{110}-H_{160}$} & \colhead{$H_{160}-K_s$} &
  \colhead{$M_{UV,AB}$\tablenotemark{b}} & \colhead{Ref\tablenotemark{c}}}
\startdata
\multicolumn{8}{c}{Candidates in the CDF-South and HDF-North Fields}\\
GNS-zD5 & 12:36:44.68 & 62:16:15.4 & 25.4$\pm$0.2 & 1.7$\pm$0.8 & 1.0$\pm$0.3 & --- & $-$21.5 & --- \\ 
GNS-zD6 & 03:32:22.66 & $-$27:43:00.6 & 25.4$\pm$0.2 & 1.6$\pm$0.5 & 0.2$\pm$0.2 & --- & $-$21.5 & [12,13] \\ 
GNS-zD4 & 12:36:10.93 & 62:09:15.6 & 25.8$\pm$0.3 & $>$1.6 & 0.7$\pm$0.3 & --- & $-$21.1 & --- \\ 
GNS-zD1 & 03:32:43.29 & $-$27:42:47.9 & 25.8$\pm$0.2 & 1.3$\pm$0.4 & 0.4$\pm$0.2 & --- & $-$21.1 & --- \\ 
HDFN-3654-1216 & 12:36:54.12 & 62:12:16.2 & 26.0$\pm$0.1 & 1.1$\pm$0.3
& 0.4$\pm$0.3 & 0.0$\pm$0.3 & $-$20.9 & [2,7] \\ 
GNS-zD3 & 03:32:06.10 & $-$27:46:37.3 & 26.1$\pm$0.3 & 1.5$\pm$1.0 & 0.8$\pm$0.5 & --- & $-$20.8 & --- \\ 
GNS-zD7\tablenotemark{d} & 03:32:42.84 & $-$27:42:47.7 & 26.1$\pm$0.2 & 1.5$\pm$1.0 & 0.4$\pm$0.5 & --- & $-$20.8 & --- \\ 
UDF-640-1417 & 03:32:42.56 &
$-$27:46:56.6 & 26.2$\pm$0.1 & 1.3$\pm$0.3 & 0.5$\pm$0.2 & 0.5$\pm$0.3
& $-$20.7 & [1,2,3,4,5,7,11,12] \\ 
GNS-zD2 & 03:32:32.03 & $-$27:45:37.2 & 26.2$\pm$0.3 & 1.6$\pm$1.0 & 0.6$\pm$0.3 & --- & $-$20.7 & --- \\ 
NICPAR2-3308-5229\tablenotemark{e} & 03:33:08.29 &
$-$27:52:29.2 & 26.7$\pm$0.1 & $>2.2$\tablenotemark{f} & 0.6$\pm$0.2 &
--- & $-$20.2 & [7,8] \\ 
CDFS-3225-4627 & 03:32:25.22 & $-$27:46:26.7 &
26.7$\pm$0.2 & 1.4$\pm$0.5 & 0.2$\pm$0.3 & $<$0.2\tablenotemark{f} & $-$20.2 & [2,7] \\ 

UDF-983-964 &
03:32:38.80 & $-$27:47:07.2 & 26.9$\pm$0.2 & $>$3.2\tablenotemark{f} &
0.1$\pm$0.3 & 0.0$\pm$0.8 & $-$20.0 & [1,2,4,5,7,8,11] \\ 

UDF-387-1125\tablenotemark{g} & 03:32:42.56 & $-$27:47:31.4 & 27.1$\pm$0.2 &
1.3$\pm$0.5 & 0.8$\pm$0.4 & $<$0.0\tablenotemark{f} & $-$19.8 & [1,5,7,11] \\

UDF-3244-4727\tablenotemark{g} & 03:32:44.02 &
$-$27:47:27.3 & 27.3$\pm$0.2 & $>$2.6\tablenotemark{f} & 0.5$\pm$0.4 &
$<$0.1\tablenotemark{f} & $-$19.6 & [7,8] \\ 

NICPAR1-3303-4111\tablenotemark{e} &
03:33:03.81 & $-$27:41:12.1 & 27.8$\pm$0.1 & $>$1.5\tablenotemark{f} &
0.4$\pm$0.2 & --- & $-$19.1 & --- \\

\multicolumn{8}{c}{Candidates in Lensing Cluster Fields\tablenotemark{a}}\\
A1689-zD1 & 13:11:29.73 & $-$01:19:20.9 &
24.7$\pm$0.1 & $>$2.2\tablenotemark{f} & 0.6$\pm$0.2 & --- & $-$19.8 & [6] \\

CL0024-zD1 & 00:26:37.93 & 17:10:39.0 & 25.6$\pm$0.1 & 1.3$\pm$0.8 & 0.4$\pm$0.3 & --- & $-$19.3 & [9] \\
CL0024-iD1 & 00:26:37.78 & 17:10:40.0 & 25.0$\pm$0.1 & 0.9$\pm$0.2 & 0.1$\pm$0.1 & --- & $-$19.9 & [9]
\enddata 
\tablenotetext{a}{There are a few weaker $z_{850}$-dropout candidates
  behind lensing clusters that have been identified by Richard et
  al.\ (2008) and Bouwens et al.\ (2009a), but none of these have deep
  enough optical data to be included in the present list as reliable
  $z\sim7$ candidates.}
\tablenotetext{b}{The absolute magnitudes estimated for the sources are for an
effective rest-frame wavelength of $\sim$1900$\AA$.  The absolute magnitudes given for the lensed sources include a correction for the estimated magnification factor (Bradley et al.\ 2008; Zheng et al.\ 2009).}
\tablenotetext{c}{References: [1] Bouwens et al.\ (2004b),
[2] Bouwens \& Illingworth (2006), [3] Yan \& Windhorst (2004), [4]
Coe et al.\ (2006), [5] Labb\'{e} et al.\ (2006), [6] Bradley et al.\
(2008), [7] Bouwens et al.\ (2008), [8] Oesch et al.\ (2009), 
[9] Zheng et al.\ (2009),
[10] Gonzalez et al.\ (2010), [11] Oesch et al.\ (2010a), 
McLure et al.\ (2010),
Bunker et al.\ (2010), Yan et al.\ (2010), Finkelstein et al.\ (2010), 
[12] Castellano et al.\ (2010), [13] Hickey et al.\ (2010)}

\tablenotetext{d}{This source has a measured $z_{850}-Y_{105}$
  color $\sim0.0\pm1.0$ and $J_{125}-H_{160}\sim0.6\pm0.2$ color in
  the high S/N WFC3/IR observations (see Table~\ref{tab:wfc3comp}).
  This suggests this candidate may be a red $z$$\sim$1-2 galaxy and not a
  $z\sim7$ star-forming galaxy.}
\tablenotetext{e}{See the footnotes in Table 2 of Bouwens et
  al.\ (2008) for these candidates}
\tablenotetext{f}{Upper and lower limits on the measured colors are
the $1\sigma$ limits.}
\tablenotetext{g}{The S/N on our $J_{110}$-band fluxes for
  UDF-387-1127 and UDF-3244-4727 is higher than initially reported in
  Bouwens et al.\ (2008).  The improved S/N is the result of 8 orbits
  of additional $J_{110}$-band observations on these sources.}
\end{deluxetable*}

\subsection{Total Sample of $z\sim7$ $z_{850}$-dropouts and $z\sim9$ $J_{110}$-dropouts}

In \S3.3 and \S3.4, we identify 7 plausible $z\sim7$ $z_{850}$-dropout
candidates in the deep, wide-area ($\sim$65 arcmin$^2$) NICMOS
$H_{160}$-band data recently obtained over the GOODS + HDF-South
fields.  Two possible $z\sim9$ $J_{110}$-dropout candidates are
identified, but appear unlikely to correspond to $z$$\sim$9 galaxies.
To increase our search area to $\sim88$ arcmin$^2$ (including all the
NICMOS observations in Table~\ref{tab:obsdata}), we combine this
sample with the sample of $z\gtrsim7$ dropouts found in the deep, but
smaller area $\sim$23 arcmin$^2$ NICMOS data already considered by
Bouwens et al.\ (2008).  Figure~\ref{fig:stampdeep} shows postage
stamp images of the $z\sim7$ field candidates from that study.  Those
postage stamps are essentially identical to those presented in Bouwens
et al.\ (2008), but incorporate 8 additional orbits of NICMOS
observations taken on two HUDF $z_{850}$-dropout candidates
(UDF-387-1127, UDF3244-4727) in the $J_{110}$ band.

Our total sample of $z\sim7$ $z_{850}$-dropouts identified in our
search fields (Table~\ref{tab:obsdata}) -- including both the old and
new data -- is presented in Table~\ref{tab:zcandlist}.  Also included
in this table are three $z\sim7$ $z_{850}$-dropouts identified in
searches behind lensing clusters (Bradley et al.\ 2008; Bouwens et
al.\ 2009; Zheng et al.\ 2009).

\subsection{Estimated Contamination Rate}

High redshift dropout selections are subject to contamination from (1)
sources that enter the selection due to photometric scatter, (2)
transient sources, (3) low-mass stars, and (4) spurious sources (i.e.,
corresponding to no real source).  We consider each of these sources
of contamination in examining the present $z\sim7$ $z_{850}$-dropout
and $z\sim9$ $J_{110}$-dropout selections.

\textit{Possible Contamination from Photometric Scatter:} In general,
the most important contaminant for high-redshift dropout selections
are sources that enter the selections due to photometric scatter
(e.g., see Bouwens et al.\ 2008; Bouwens et al.\ 2010b).  We estimate
the contamination rate that results from this effect by starting with
the color distribution observed for $H$$\sim$25.0-26.5 sources in the
HUDF09 WFC3/IR field (Bouwens et al.\ 2010b; Oesch et al.\ 2010a) and
then adding photometric scatter to match that expected for each source
in our NICMOS catalogs.  Repeating these simulations for each source
in our NICMOS catalogs (from all search fields not considered in
Bouwens et al. 2008), we predict that 1.7 sources and 0.2 sources
would enter our $z_{850}$-dropout and $J_{110}$-dropout selections,
respectively, via photometric scatter.  The implied contamination rate
is 24\% and 10\%, respectively, which is somewhat higher than
estimates for our HUDF $z_{850}$-dropout selections (Bouwens et
al.\ 2008) or for our new WFC3/IR results (Oesch et al.\ 2010a;
Bouwens et al.\ 2010b).  This is due to the somewhat shallower depths
of the ACS optical data over the GOODS fields and larger prevalence of
low-redshift sources with similar colors to high-redshift galaxies at
bright magnitudes (see also \S3.3).  The above procedure is
essentially identical to that used in Bouwens et al.\ (2008).

\textit{Possible Contamination from Transient Sources:} Contamination
from transient sources, particularly SNe, are potentially important
for each of our search fields, given that the NICMOS observations were
typically acquired at least 4 years after the ACS optical
observations.  Using the GOODS SNe searches (Riess et al.\ 2004;
Strolger et al.\ 2004) as a baseline, Bouwens et al.\ (2008) argued
that the contamination rate from SNe should be no larger than $0.012\,
\textrm{arcmin}^{-2}$ for observations where the optical and near-IR
$J+H$ observations are taken at very different times.  This suggests a
contamination rate of $\sim$1 source for the present search for
$z\gtrsim7$ galaxies over $\sim$65 arcmin$^2$ of new data.  However,
the above calculation assumes the $J$ and $H$ band observations are
acquired at the same time.  In reality, for most of the new data, the
$J_{110}$-band observations were taken at least 3-12 months after the
NICMOS $H_{160}$-band observations for each of our new
$z_{850}$-dropout candidates.  Consequently, the measured $J-H$ colors
for SNe's found in GOODS NICMOS survey would likely be very red and
hence show up as an apparent $J_{110}$-dropout candidate.  We might
therefore expect 1 SNe contaminants to be present in our
$J_{110}$-dropout selection.  This may be the case for GNS-JD2 (see
\S3.4).

\textit{Possible Contamination from Low-mass Stars:} Low-mass stars
(e.g., T or L dwarfs) have very similar $z-J$, $J-H$ colors to
star-forming galaxies at $z\sim7$ and therefore could contaminate our
$z_{850}$-dropout selection.  Given the modest resolution of the
NICMOS data, it is difficult to use these data to clearly distinguish
small ($\sim$0.15$''$ half-light radius) star-forming galaxies found
at $z\sim7$ from unresolved low-mass stars.  Fortunately, the
$z_{850}$-band observations have sufficient resolution and depth that
low-mass stars (with $z-J$ colors $\sim$1.5-2) would be evident at
$5\sigma$ (0.2$''$-diameter aperture) in the $z_{850}$-band images as
point sources.  Only one possible T dwarf candidates is evident in the
new wide-area NICMOS observations and that is the source at
03:32:22.66, $-$27:43:00.6 (GNS-zD6).  However, that source appears to
be resolved in the new WFC3/IR observations as part of the Early
Release Science Program (GO11359: PI O'Connell), so that source
appears unlikely to be a low mass star.  The present situation is
somewhat in contrast to the NICMOS observations considered by Bouwens
et al.\ (2008) where two probable T-dwarfs were identified in
selecting $z\gtrsim7$ galaxies (included on Figure~\ref{fig:zjjh} as
the red triangles).

\textit{Possible Contamination from Spurious Sources:} Contamination
from spurious sources could be a concern for our $z\sim9$
$J_{110}$-dropout selection.  Each of our $J_{110}$-dropout candidates
is only detected in a single band and the estimated $\sim$6$\sigma$
significance of these detections may be an overestimate (due to real
data possessing many non-Gaussian characteristics).  Therefore, to
estimate the likely number of spurious sources, we used the standard
negative image test (e.g., Dickinson et al.\ 2004) and repeated our
$z\sim9$ $J_{110}$-dropout selection on the negative $H_{160}$-band
images (after masking out sources brighter than 23 mag).  Such a
selection yielded $\sim$12 $J_{125}$-dropout candidates, but all of
the candidates (on the negative images) could be eliminated due to an
obvious association with defects or irregularities in the reduction.
While this might suggest that spurious sources do not dominate our
$J_{110}$-dropout selection, there were a modest number of formal
$5\sigma$ detections on the negative images quite close to satisfying
our $J_{110}$-dropout selection criteria, and so contamination from
spurious sources here is certainly not impossible.

By contrast, for our $z\sim7$ $z_{850}$-dropout selection, spurious
sources are not an important concern.  Each of our candidates is
detected at $\gtrsim$2$\sigma$ in the $J_{110}$ band and
$\gtrsim$5$\sigma$ in the $H_{160}$ band -- making contamination from
spurious sources extremely unlikely.

\textit{Summary:} In total, we expect 1.7 and 1.2 contaminants in our
$z\sim7$ $z_{850}$-dropout and $z\sim9$ $J_{110}$-dropout selections,
respectively (equivalent to contamination levels of 24\% and 60\%).
The only meaningful source of contamination for our $z_{850}$-dropout
selections is photometric scatter, while for our $J_{110}$-dropout
selection, several sources of contamination contribute.  We expect
$\sim$1 contaminant from transient sources (SNe), 0.2 contaminants
from photometric scatter, and $\lesssim$1 contaminants from spurious
sources.

\begin{figure*}
\epsscale{1.15}
\plottwo{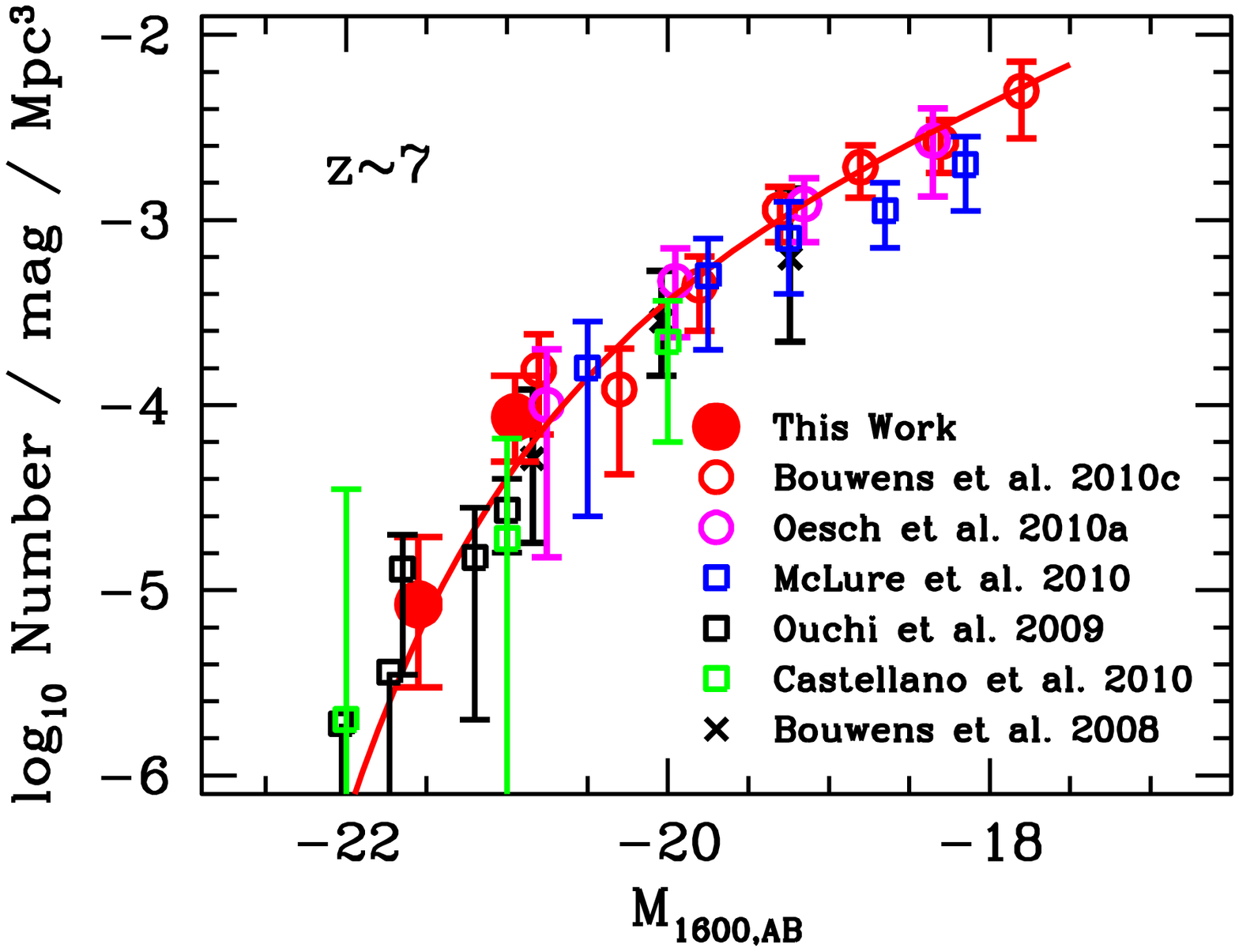}{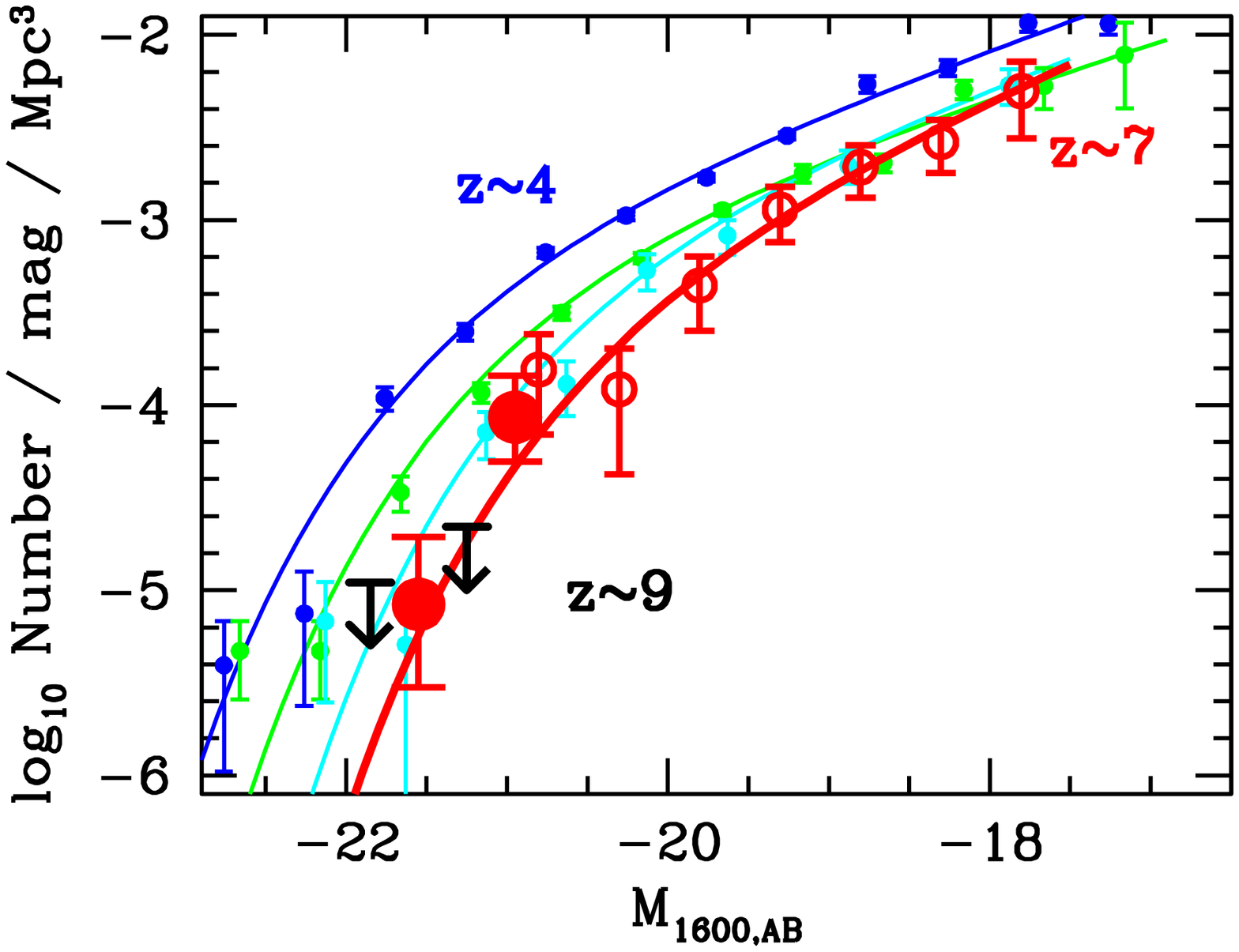}
\caption{(\textit{left}) Present determination of the stepwise $UV$ LF
  at $z\sim7$ using deep wide-area NICMOS + ground-based observations
  (\textit{large solid red circles}: see \S4).  The $z\sim7$ $UV$ LF
  is only redetermined here at luminosities where we can take
  advantage of the new $\sim$65 arcmin$^2$ of NICMOS observations (see
  Table~\ref{tab:obsdata}) to search for candidate $z\gtrsim7$
  galaxies.  For comparison, we also show the $z\sim7$ LFs reported by
  Bouwens et al.\ (2008: \textit{black crosses}), Ouchi et al.\ (2009:
  \textit{black squares}), Castellano et al.\ (2009: \textit{green
    squares}), Oesch et al.\ (2010a: \textit{open magenta circles}),
  McLure et al.\ (2010: \textit{blue squares}), and Bouwens et
  al.\ (2010c: \textit{open red circles}).  Constraints from Oesch et
  al.\ (2009) and Wilkins et al.\ (2010b) are similar, but not shown
  to reduce confusion.  The solid red line is the best-fit Schechter
  function presented in Oesch et al.\ (2010a).  (\textit{right})
  Stepwise $UV$ LF determined here at $z\sim7$ (\textit{solid red
    circles}) versus that derived at $z\sim4$ (\textit{blue}),
  $z\sim5$ (\textit{green}), and $z\sim6$ (\textit{cyan}) by Bouwens
  et al.\ (2007).  The open red circles and red line represent the
  $z\sim7$ LF determined by Bouwens et al.\ (2010c).  Also shown are
  our constraints on the LF at $z$$\sim$9 from the present
  $J_{110}$-dropout search (\textit{black upper limits}: see \S4).
  Similar to the situation for our $z\sim7$ LF, the $z\sim9$
  $J_{110}$-dropout LF is only computed at luminosities where we can
  take advantage of the new NICMOS data.  The volume density of
  $\sim$$L_{z=3}^{*}$ ($\sim$$-$21 mag) star-forming galaxies at
  $z\sim7$ and $z\sim9$ is 13$_{-5}^{+8}$$\times$ and $>$25$\times$
  lower, respectively, than at $z\sim4$.\label{fig:lf}}
\end{figure*}

\section{Implied Constraints on the Rest-frame UV Luminosity Function}

The present wide-area ($\sim$88 arcmin$^2$) NICMOS search for
$z\gtrsim7$ candidates provides a useful constraint on the volume
density of luminous star-forming galaxies at $z\sim7$ and $z\sim9$.
These observations cover twice as much area as $\sim$39 arcmin$^2$ ERS
observations and cover similar area to that available from the
CDF-South HAWK-I observations (90 arcmin$^2$: Castellano et al.\ 2010;
Hickey et al.\ 2010).  They therefore provide a somewhat comparable
constraint on the bright end (i.e., $M_{UV}<-20.7$) of the $z\gtrsim7$
LFs to these studies.  And while the Ouchi et al.\ (2009) Subaru
Suprime-Cam search extends over considerably more area (1568
arcmin$^2$) than these observations, the present NICMOS search is
somewhat deeper than that study (by $\sim$0.7 mag).  The present
search also benefits from somewhat higher quality multiwavelength data
-- deep high-resolution ACS or mid-IR IRAC data (Giavalisco et
al.\ 2004; Dickinson \& GOODS Team 2004) -- than are generally
  available for the Subaru fields.  This allows for a more robust
  discrimination between high-redshift star-forming galaxies and
  low-redshift contaminants (and low mass stars).

Here we will estimate the stepwise LF $\Phi(M)$ using the relatively
simple $V/V_{eff}$ approach (e.g., Steidel et al.\ 1999) where
\begin{equation}
\Phi(m) = \frac{N(m)}{V_{eff}(m)}
\end{equation}
where $N(m)$ is the number of $z\gtrsim7$ candidates in a given
magnitude interval $m$ (from all of our search fields) and $V_{eff}
(m)$ is the effective volume in the magnitude interval $m$.

To compute the effective volume with which we can select star-forming
galaxies in our various search fields, as a function of magnitude and
redshift, we run detailed Monte-Carlo simulations where we introduce
model galaxies into the observed data and attempt to select them as
dropouts using the procedures laid out in \S3.  These simulations are
performed in the same way as they were performed in Bouwens et
al.\ (2008).  For the model galaxies in these simulations, we assume
that the $z\gtrsim7$ galaxies of a given luminosity have similar
pixel-by-pixel morphologies to $z\sim4$ galaxies in the HUDF with the
same luminosity, but scaled in size as $(1+z)^{-1}$ to match the
observed size-redshift trends at $z\geq2$ (e.g., Ferguson et
al.\ 2004; Bouwens et al.\ 2004; Bouwens et al.\ 2006; Oesch et
al.\ 2010b).  The $UV$-continuum slopes of the model galaxies are
assumed to have a mean of $-2$, with a $1\sigma$ scatter of 0.5.
These latter assumptions match the colors of bright star-forming
galaxies found at $z\sim7$ (Bouwens et al.\ 2008; Oesch et al.\ 2010a;
Bunker et al.\ 2010; Bouwens et al.\ 2010b; Finkelstein et al.\ 2010).

\begin{deluxetable}{lcc}
\tablewidth{0pt}
\tabletypesize{\footnotesize}
\tablecaption{Stepwise Constraints on the rest-frame $UV$ LF at $z\sim7$ and $z\sim9$ from Wide-Area NICMOS + Ground-Based Observations (\S4).\tablenotemark{a}\label{tab:swlf}}
\tablehead{
\colhead{$M_{UV,AB}$\tablenotemark{b}} & \colhead{$\phi_k$ ($10^{-6}$ Mpc$^{-3}$ mag$^{-1}$)}}
\startdata
\multicolumn{2}{c}{$z_{850}$-dropouts ($z\sim7$)}\\
$-$21.55 & $8_{-5}^{+11}$ \\
$-$20.95 & $86_{-37}^{+58}$ \\
\multicolumn{2}{c}{$J$-dropouts ($z\sim9$)}\\
$-$21.85 & $<11$\tablenotemark{c}\\
$-$21.25 & $<22$\tablenotemark{c}\\
\enddata 
\tablenotetext{a}{The $UV$ LFs are only redetermined here at
  luminosities where we can take advantage of the new $\sim$65
  arcmin$^2$ of NICMOS observations (see Table~\ref{tab:obsdata}) to
  search for candidate $z\gtrsim7$ galaxies.}
\tablenotetext{b}{The effective rest-frame wavelength is
  $\sim$1900\AA$\,$ for our $z_{850}$-dropout selection and
  $\sim$1500\AA$\,$ for our $J_{110}$-dropout selection.}
\tablenotetext{c}{Upper limits here are $1\sigma$ (68\% confidence).}
\end{deluxetable}

In computing $UV$ LFs on the basis of our $z\gtrsim7$ search, we
consider all seven $z\sim7$ $z_{850}$-dropout candidates in the newest
NICMOS data and suppose that $\sim$5.3 $z_{850}$-dropout candidates
from this search correspond to $z\sim7$ galaxies (accounting for the
24\% contamination rate estimated in \S3.6).  We also incorporate the
constraints from the NICMOS + ground-based $z\sim7$ search considered
by Bouwens et al.\ (2008).  In addition, we suppose that $\lesssim$1
$J_{110}$-dropout candidates from this search correspond to $z$$\sim$9
galaxies, given the concerns that exist for each candidate (\S3.4).
Finally, the $z\sim7$ and $z\sim9$ LFs are only redetermined here at
luminosities where we can take advantage of the new $\sim$65
arcmin$^2$ of NICMOS observations (see Table~\ref{tab:obsdata}).

Our stepwise $UV$ LF at $z\sim7$ is presented in Table~\ref{tab:swlf}
and in Figure~\ref{fig:lf} (\textit{left}).  For comparison,
Figure~\ref{fig:lf} also includes the $z\sim7$ LFs of Bouwens et
al.\ (2008), Ouchi et al.\ (2009), Castellano et al.\ (2010), Oesch et
al.\ (2010a), McLure et al.\ (2010), and Bouwens et al.\ (2010c).  The
present LF results are in reasonable agreement with previous
determinations.  In the right panel of Figure~\ref{fig:lf}, the
present $z\sim7$ LF results are shown relative to the LFs at $z\sim4$,
$z\sim5$, $z\sim6$ (from Bouwens et al.\ 2007) to provide a sense of
the evolution from $z\sim4$.  Also included on this figure
(\textit{black upper limits}) are the constraints on the $UV$ LF at
$z\sim9$ from the present $J_{110}$-dropout search.  The $UV$ LF is
13$_{-5}^{+8}$$\times$ lower at $z\sim7$ than at $z\sim4$ and
$>$25$\times$ lower ($1\sigma$) at $z\sim9$ than at $z\sim4$.
\textit{The latter constraint is the most stringent constraint yet
  available on the volume density of $\gtrsim L_{z=3}^{*}$ galaxies at
  $z\sim9$.}

Of course, the above LF determinations are subject to uncertainties as
a result of large-scale structure (``cosmic variance''), and therefore
to properly frame the above results, it is helpful to estimate the
size of these uncertainties.  For convenience, we utilize the Trenti
\& Stiavelli (2008) cosmic variance calculator to make this estimate.
Given that the approximate volume density of the sources probed at
$z\sim7$ and $z\sim9$ is $\sim$6$\times$10$^{-5}$ Mpc$^{-3}$ and
$\sim$2$\times$10$^{-5}$ Mpc$^{-3}$, respectively, the approximate
bias parameters are 10 and 12, respectively (e.g., Somerville et
al.\ 2004; Trenti \& Stiavelli et al.\ 2008).  Given that our NICMOS
search data are fairly randomly scattered over the two GOODS fields,
we assume two independent search fields of dimension $10'\times16'$
and adopt a width for our redshift selection window $\Delta z\sim1.5$
(e.g., see Figure 7 of Bouwens et al.\ 2008).  Utilizing these inputs,
we estimate that the large-scale structure uncertainties on our LF
results here are $\sim$17\% and $\sim$20\%, respectively.  Large-scale
structure uncertainties are therefore much smaller than those
uncertainties estimated from the small number statistics ($\sim$8
sources) and uncertain contamination rates (\S3.6).

\section{Summary}

We have taken advantage of $\sim$88 arcmin$^2$ of deep, wide-area
NICMOS data to search for $z\gtrsim7$ galaxies within the HUDF, the
two GOODS fields, and the HDF South.  This search incorporates
$\sim$65 arcmin$^2$ of wide-area NICMOS data not previously used to
identify $z\gtrsim7$ galaxies.\footnote{A small fraction of these 65
  arcmin$^2$ (4 arcmin$^2$) had previously been used by Henry et
  al.\ (2009) for a $z\sim7$ $z_{850}$-dropout search.}  $\sim$248
arcmin$^2$ of deep ground-based data ($\gtrsim25.5$ mag, $5\sigma$)
previously used by Bouwens et al.\ (2008) is also considered.  In
total, we find $\sim$7 plausible $z\sim7$ $z_{850}$-dropout candidates
in the new NICMOS observations (six of which are being reported for
the first time) and $\sim$2 possible (but probably unlikely) $z\sim9$
$J_{110}$-dropout candidates.  These candidates significantly add to
the number of luminous $z\gtrsim7$ candidates known within the GOODS
fields and improve our constraints on the volume density of luminous
galaxies at $z\sim7$.  These candidates have recently been used to
model the stellar populations of bright $z\sim7$ galaxies (Gonzalez et
al.\ 2010; Labb{\'e} et al.\ 2010b).

When taken together with the NICMOS data already considered in Bouwens
et al.\ (2008), we have identified 15 $z\sim7$ $z_{850}$-dropout
candidates in total from $\sim$88 arcmin$^2$ of deep NICMOS data.
After running detailed simulations to estimate the selection volumes
and contamination rates, we use our new $z\sim7$ samples (plus
ground-based search area) to update the Bouwens et al.\ (2008)
determination of the $UV$ LF at $z\sim7$ and to strengthen our
constraints on the LF at $z\sim9$.

We find that the bright end of the UV LF at $z\sim7$ is
13$_{-5}^{+8}$$\times$ lower at $z\sim7$ than at $z\sim4$.  At
$z\sim9$, the $UV$ LF is a factor of $>$25$\times$ lower ($1\sigma$)
than at $z\sim4$, assuming that at most one of the $J_{110}$-dropout
candidates identified here is at $z\sim9$.

\acknowledgements

We are grateful to our HST program coordinator Beth Perrillo and
contact scientist Luigi Bedin for the substantial assistance they
provided in setting up the follow-up observations (GO11144) for all of
the $z\gtrsim7$ candidates we identified in deep NICMOS $H_{160}$-band
data.  We are thankful to Louis Bergeron, Susan Kassin, and Rodger
Thompson for their help in reducing some of the NICMOS data used in
this study.  This work also benefitted from stimulating conversations
with Romeel \dave, Claude-Andr{\'e} Faucher-Gigu{\`e}re, Kristian
Finlator, Cedric Lacey, Pascal Oesch, Masami Ouchi, and Piero Rosati.
The comments of the anonymous referee significantly improved the
clarity of the paper.  We appreciate Pascal Oesch's giving this paper
a careful reading.  We acknowledge support from NASA grants
HST-GO09803.05-A, HST-GO10937.03-A, HST-GO11082.04-A, HST-GO11144.03-A
and NAG5-7697.

\begin{deluxetable}{ccccccc}
\tablewidth{0pt}
\tabletypesize{\footnotesize}
\tablecaption{Measurements of the fluxes of our $z\gtrsim7$ candidates in the optical ACS and near-IR NICMOS data.\tablenotemark{a}\label{tab:fluxdata}}
\tablehead{
\colhead{} & \multicolumn{6}{c}{$f_{\nu}$ (nJy)} \\
\colhead{Object ID} & \colhead{$B_{435}$} & \colhead{$V_{606}$} & \colhead{$i_{775}$} & \colhead{$z_{850}$} & \colhead{$J_{110}$} & \colhead{$H_{160}$}}
\startdata
GNS-zD1 & $-$10$\pm$9 & 0$\pm$6 & 15$\pm$12 & 35$\pm$14 & 124$\pm$13 & 169$\pm$21\\
GNS-zD2 & 0$\pm$9 & 9$\pm$7 & $-$8$\pm$13 & 12$\pm$13 & 69$\pm$14 & 116$\pm$21\\
GNS-zD3 & $-$2$\pm$8 & 6$\pm$6 & $-$5$\pm$9 & 12$\pm$13 & 64$\pm$29 & 130$\pm$25\\
GNS-zD4 & 5$\pm$17 & 9$\pm$13 & $-$14$\pm$22 & $-$9$\pm$23 & 102$\pm$24 & 201$\pm$32\\
GNS-zD5 & $-$2$\pm$13 & $-$1$\pm$9 & $-$29$\pm$14 & 23$\pm$17 & 105$\pm$22 & 258$\pm$28\\
GNS-zD6 & 3$\pm$14 & $-$8$\pm$10 & 33$\pm$17 & 46$\pm$21 & 208$\pm$29 & 258$\pm$25\\
GNS-zD7 & 9$\pm$13 & $-$15$\pm$9 & $-$3$\pm$16 & 21$\pm$18 & 93$\pm$17 & 128$\pm$26\\
GNS-JD1 & --- & 7$\pm$8 & 20$\pm$16 & $-$29$\pm$26 & 53$\pm$17 & 140$\pm$22\\
GNS-JD2 & 1$\pm$8 & 3$\pm$7 & 9$\pm$8 & 16$\pm$9 & $-$20$\pm$16 & 166$\pm$26\\
\enddata 
\tablenotetext{a}{Uncertainties here are $1\sigma$.  $m_{AB} = 31.4 - 2.5\log_{10}(f_{\nu} \textrm{[nJy]})$.}
\end{deluxetable}

\appendix

\section{Flux Measurements for our Individual $z\gtrsim7$ Candidates}

To better assess the significance of the colors measured for our
$z\gtrsim7$ candidates, we also tabulate the observed fluxes of each
candidate in Table~\ref{tab:fluxdata}.  The tabulated fluxes also
allow us to readily evaluate whether any of the candidates is detected
in the optical (and therefore not likely at high redshift).


\begin{thebibliography}{} 
\bibitem[Beckwith et al.(2006)]{2006AJ....132.1729B} Beckwith, S.~V.~W., et 
al.\ 2006, \aj, 132, 1729
\bibitem[Bertin and Arnouts (1996)]{1996A&AS..117..393B} Bertin, E.\ and 
Arnouts, S.\ 1996, \aaps, 117, 39
\bibitem[Bouwens et al.(2004)]{2004ApJ...611L...1B} Bouwens, R.~J., 
Illingworth, G.~D., Blakeslee, J.~P., Broadhurst, T.~J., 
\& Franx, M.\ 2004a, \apjl, 611, L1
\bibitem[Bouwens et al.\ (2004)]{2004} Bouwens, R.~J., et al.\  2004b, \apjl,
616, L79
\bibitem[Bouwens et al. (2006)]{2006Bouwens} Bouwens, R.J., Illingworth,
G.D., Blakeslee, J.P., \& Franx, M.  2006, \apj, 653, 53 
\bibitem[Bouwens \& Illingworth(2006)]{2006Natur.443..189B} Bouwens, R.~J., 
\& Illingworth, G.~D.\ 2006, \nat, 443, 189
\bibitem[Bouwens et al.(2007)]{2007ApJ...670..928B} Bouwens, R.~J., 
Illingworth, G.~D., Franx, M., \& Ford, H.\ 2007, \apj, 670, 928
\bibitem[Bouwens et al.\ (2008)]{2008ApJ...686..230B} Bouwens, R.~J., 
Illingworth, G.~D., Franx, M., \& Ford, H.\ 2008, \apj, 686, 230 
\bibitem[Bouwens et al.(2009)]{2009ApJ...690.1764B} Bouwens, R.~J., et al.\ 
2009a, \apj, 690, 1764
\bibitem[Bouwens et al.\ (2009)]{2009ApJ...705..936B} Bouwens, R.~J., et
  al.\ 2009b, \apj, 705, 936
\bibitem[Bouwens et al. (2010a)]{eee2} Bouwens, R.J., et al.\ 2010a,
  \apj, 708, L69
\bibitem[Bouwens et al. (2010b)]{eee2} Bouwens, R.J., et al.\ 2010b,
  \apj, 709, L133
\bibitem[Bouwens et al.(2010)]{2010arXiv1006.4360B} Bouwens, R.~J., et
  al.\ 2010c, \apj, submitted, arXiv:1006.4360
\bibitem[Bradley et al. (2008)]{Bradley} Bradley, L.D., et al. 2008, \apj, 678, 647
\bibitem[Brandt et al.(2001)]{2001AJ....122....1B} Brandt, W.~N., et al.\ 
2001, \aj, 122, 1
\bibitem[Bunker et al. (2010)]{eee} Bunker, A., et al.\ 2010, \mnras,
  in press, arXiv:0909.2255
\bibitem[Castellano et 
al.(2010)]{2010A&A...511A..20C} Castellano, M., et al.\ 2010, \aap, 511, A20
\bibitem[Coe et al.(2006)]{2006AJ....132..926C} Coe, D., Ben{\'{\i}}tez, 
N., S{\'a}nchez, S.~F., Jee, M., Bouwens, R., \& Ford, H.\ 2006, \aj, 132, 
926 
\bibitem[Conselice et al.(2010)]{2010arXiv1010.1164C} Conselice, C.~J., et 
al.\ 2010, \mnras, submitted, arXiv:1010.1164 
\bibitem[Dickinson(1998)]{1998hdf..symp..219D} Dickinson, M.\ 1998, The 
Hubble Deep Field, 219
\bibitem[Dickinson et al.(2004)]{2004ApJ...600L..99D} Dickinson, M.~et al.\ 
2004, \apjl, 600, L99
\bibitem[Dickinson 
\& GOODS(2004)]{2004AAS...20516301D} Dickinson, M., \& GOODS 2004, Bulletin of the American Astronomical Society, 36, 1614 
\bibitem[Ferguson et al.(2004)]{2004ApJ...600L.107F} Ferguson, H.~C.~et 
al.\ 2004, \apjl, 600, L107
\bibitem[Finkelstein et al.(2010)]{2010ApJ...719.1250F} Finkelstein, S.~L., 
Papovich, C., Giavalisco, M., Reddy, N.~A., Ferguson, H.~C., Koekemoer, 
A.~M., \& Dickinson, M.\ 2010, \apj, 719, 1250
\bibitem[Giavalisco et al.(2004)]{2004ApJ...600L..93G} Giavalisco, M., et 
al.\ 2004, \apjl, 600, L93
\bibitem[Gonz{\'a}lez et al.(2010)]{2010ApJ...713..115G} Gonz{\'a}lez, V., 
Labb{\'e}, I., Bouwens, R.~J., Illingworth, G., Franx, M., Kriek, M., 
\& Brammer, G.~B.\ 2010, \apj, 713, 115
\bibitem[Henry et al.(2009)]{2009ApJ...697.1128H} Henry, A.~L., et al.\ 
2009, \apj, 697, 1128
\bibitem[Hickey et al.(2010)]{2010MNRAS.404..212H} Hickey, S., Bunker, A., 
Jarvis, M.~J., Chiu, K., \& Bonfield, D.\ 2010, \mnras, 404, 212 
\bibitem[Kajisawa et al.(2006)]{2006PASJ...58..951K} Kajisawa, M., et al.\ 
2006, \pasj, 58, 951 
\bibitem[Kimble et al.(2006)]{2006SPIE.6265E..14K} Kimble, R.~A., MacKenty, 
J.~W., \& O'Connell, R.~W.\ 2006, \procspie, 6265, 14
\bibitem[Komatsu et al.(2010)]{2010arXiv1001.4538K} Komatsu, E., et al.\ 
2010, \apj, submitted, arXiv:1001.4538 
\bibitem[Kron(1980)]{1980ApJS...43..305K} Kron, R.\ G.\ 1980, \apjs, 43, 
305
\bibitem[Labb{\'e} et al.(2003)]{2003AJ....125.1107L} Labb{\'e}, I., et 
al.\ 2003, \aj, 125, 1107 
\bibitem[labbe et al.(2006)]{2006astro.ph..8444L} Labb\'{e}, I., Bouwens, R., 
Illingworth, G.~D., \& Franx, M.\ 2006, \apjl, 649, 67
\bibitem[Labb{\'e} et al.(2010)]{2010ApJ...708L..26L} Labb{\'e}, I., et 
al.\ 2010a, \apjl, 708, L26
\bibitem[Labbe et al.(2010)]{2009arXiv0911.1356L} Labb{\'e}, I., et
  al.\ 2010b, \apjl, 716, L103
\bibitem[Magee et al.(2007)]{2007ASPC..376..261M} Magee, D.~K.,
  Bouwens, R.~J., \& Illingworth, G.~D.\ 2007, Astronomical Data
  Analysis Software and Systems XVI, 376, 261
\bibitem[Mannucci et al. (2006)]{mannucci} Mannucci, F., Buttery, H.,
Maiolino, R., Marconi, A. \& Pozzetti, L. 2006, \aap, 461, 423
\bibitem[McLure et al.(2010)]{2010MNRAS.403..960M} McLure, R.~J., Dunlop, 
J.~S., Cirasuolo, M., Koekemoer, A.~M., Sabbi, E., Stark, D.~P., Targett, 
T.~A., \& Ellis, R.~S.\ 2010, \mnras, 403, 960
\bibitem[Mu{\~n}oz 
\& Loeb(2008)]{2008MNRAS.386.2323M} Mu{\~n}oz, J.~A., \& Loeb, A.\ 2008, \mnras, 386, 2323 
\bibitem[Oesch et al.(2007)]{2007ApJ...671.1212O} Oesch, P.~A., et al.\ 
2007, \apj, 671, 1212 
\bibitem[Oesch et al.(2009)]{2009ApJ...690.1350O} Oesch, P.~A., et al.\ 
2009, \apj, 690, 1350
\bibitem[Oesch et al. (2010)]{eee} Oesch, P.A., et al.\ 2010a, \apj,
709, L16
\bibitem[Oesch et al. (2010b)]{eeef} Oesch, P.A., et al.\ 2010b, \apj,
709, L21
\bibitem[Oke \& Gunn(1983)]{1983ApJ...266..713O} Oke, J.~B., \& Gunn, 
J.~E.\ 1983, \apj, 266, 713 
\bibitem[Ouchi et al.(2007)]{2007ASPC..379...47O} Ouchi, M., Tokoku,
C., Shimasaku, K., \& Ichikawa, T.\ 2007, Astronomical Society of the
Pacific Conference Series, 379, 47
\bibitem[Ouchi et al.(2009)]{2009ApJ...706.1136O} Ouchi, M., et al.\ 2009, 
\apj, 706, 1136
\bibitem[Reddy \& Steidel(2009)]{2009ApJ...692..778R} Reddy, N.~A., \&
  Steidel, C.~C.\ 2009, \apj, 692, 778
\bibitem[Retzlaff et al.(2010)]{2010A&A...511A..50R} Retzlaff, J.,
  Rosati, P., Dickinson, M., Vandame, B., Rit{\'e}, C., Nonino, M.,
  Cesarsky, C., \& GOODS Team 2010, \aap, 511, A50
\bibitem[Richard et al.(2008)]{2008ApJ...685..705R} Richard, J., Stark, 
D.~P., Ellis, R.~S., George, M.~R., Egami, E., Kneib, J.-P., 
\& Smith, G.~P.\ 2008, \apj, 685, 705 
\bibitem[Riess et al.(2004)]{2004ApJ...607..665R} Riess, A.~G., et al.\ 
2004, \apj, 607, 665 
\bibitem[Riess et al.(2007)]{2007ApJ...659...98R} Riess, A.~G., et al.\ 
2007, \apj, 659, 98
\bibitem[Rosati et al.(2002)]{2002ApJ...566..667R} Rosati, P., et al.\ 
2002, \apj, 566, 667
\bibitem[Siana et al.(2007)]{2007ApJ...668...62S} Siana, B., et al.\ 2007, \apj, 668, 62
\bibitem[Siana et al.(2010)]{2010arXiv1001.3412S} Siana, B., et al.\ 2010, \apj, submitted, arXiv:1001.3412
\bibitem[Somerville et al.(2004)]{2004ApJ...600L.171S} Somerville, R.~S., 
Lee, K., Ferguson, H.~C., Gardner, J.~P., Moustakas, L.~A., \& Giavalisco, 
M.\ 2004, \apjl, 600, L171
\bibitem[Stanway et al.(2008)]{2008MNRAS.386..370S} Stanway, E.~R., Bremer, 
M.~N., Squitieri, V., Douglas, L.~S., 
\& Lehnert, M.~D.\ 2008, \mnras, 386, 370
\bibitem[Steidel et al.\ (1999)]{1999ApJ...519....1S} Steidel, C.\ C.,
Adelberger, K.\ L., Giavalisco, M., Dickinson, M.\ and Pettini, M.\ 1999,
\apj, 519, 1
\bibitem[Strolger et al.(2004)]{2004ApJ...613..200S} Strolger, L.-G., et 
al.\ 2004, \apj, 613, 200 
\bibitem[Szalay et al.(1999)]{1999AJ....117...68S} Szalay, A.~S.,
Connolly, A.~J., \& Szokoly, G.~P.\ 1999, \aj, 117, 68
\bibitem[Thompson et al.(1999)]{1999AJ....117...17T} Thompson, R.~I., 
Storrie-Lombardi, L.~J., Weymann, R.~J., Rieke, M.~J., Schneider, G., 
Stobie, E., \& Lytle, D.\ 1999, \aj, 117, 17 
\bibitem[Thompson et al.(2005)]{2005AJ....130....1T} Thompson, R.~I., et 
al.\ 2005, \aj, 130, 1
\bibitem[Trenti \& Stiavelli(2008)]{2008ApJ...676..767T} Trenti, M.,
  \& Stiavelli, M.\ 2008, \apj, 676, 767
\bibitem[Wilkins et al.(2010)]{2010MNRAS.403..938W} Wilkins, S.~M., Bunker, 
A.~J., Ellis, R.~S., Stark, D., Stanway, E.~R., Chiu, K., Lorenzoni, S., 
\& Jarvis, M.~J.\ 2010a, \mnras, 403, 938
\bibitem[Wilkins et al.(2010)]{2010arXiv1002.4866W} Wilkins, S.~M., Bunker, 
A.~J., Lorenzoni, S., \& Caruana, J.  2010b, \mnras, in press, arXiv:1002.4866 

\bibitem[Williams, R.E, et al.\ 1996]{wil96} Williams, R.E., et al.\
1996, \aj, 112, 1335.
\bibitem[Williams et al.(2000)]{2000AJ....120.2735W} Williams, R.~E., et 
al.\ 2000, \aj, 120, 2735
\bibitem[Yan \& Windhorst(2004)]{yan4} Yan, H.~\& Windhorst, R.~A.\ 2004, \apjl, 612, L93
\bibitem[Yan et al.(2010)]{2010RAA....10..867Y} Yan, H.-J., Windhorst, 
R.~A., Hathi, N.~P., Cohen, S.~H., Ryan, R.~E., O'Connell, R.~W., 
\& McCarthy, P.~J.\ 2010, Research in Astronomy and Astrophysics, 10, 867 
\bibitem[Zheng et al.(2009)]{2009ApJ...697.1907Z} Zheng, W., et
  al.\ 2009, \apj, 697, 1907
\bibitem[Zirm et al.(2007)]{2007ApJ...656...66Z} Zirm, A.~W., et al.\ 2007, 
\apj, 656, 66
\end{thebibliography}
\end{document}